\documentclass[11pt,a4paper]{article}

\usepackage{booktabs}
\usepackage{jheppub}
\usepackage{capt-of}
\usepackage{comment}
\usepackage{siunitx}
\usepackage{amssymb}
\usepackage{enumitem}   
\usepackage[english]{babel}
\usepackage{url}
\usepackage{mathtools}
\usepackage{soul}
\usepackage{bbold}
\usepackage{slashed}
\usepackage{multirow}
\usepackage{slashed}
\usepackage{multirow}
\usepackage[usenames,dvipsnames,svgnames]{xcolor}
\usepackage{appendix}
\usepackage{braket}
\usepackage{graphicx}
\usepackage{subcaption}
\usepackage[normalem]{ulem}

\begin{document}



\title{Metastable strings at PTAs: classical stability analysis}

\author[a]{Simone Blasi,}
\author[b]{Maxime Grandjean,}
\author[b]{Alberto Mariotti}

\affiliation[a]{Deutsches Elektronen-Synchrotron DESY, Notkestr. 85, 22607 Hamburg, Germany}
\affiliation[b]{Theoretische Natuurkunde and IIHE/ELEM, Vrije Universiteit Brussel, \& The International
Solvay Institutes, Pleinlaan 2, B-1050 Brussels, Belgium}

\emailAdd{simone.blasi@desy.de}
\emailAdd{Maxime.Achille.G.Grandjean@vub.be}
\emailAdd{Alberto.Mariotti@vub.be}


\abstract{ 
Metastable strings can arise from a two-step symmetry breaking chain of the type $SU(2) \to U(1) \to 1$.
They can decay through quantum tunneling by nucleating a monopole-antimonopole pair, and are prominent candidates for explaining the gravitational wave background detected at Pulsar Timing Arrays (PTAs).
We investigate the classical stability of the strings arising in this commonly-considered setup, which serves as a fundamental input for discussing their possible decay channels. 
We identify the regions of parameter space in which the strings are either classically stable or unstable. Our results show that classical instabilities can impact the parameter space relevant for PTAs.
We also discuss 
 the possible fate of the string network in the regions of classical instability.
}

\preprint{
\begin{flushright}
DESY-26-062
\end{flushright}
}

\maketitle

\section{Introduction}
Topological defects are fascinating objects that could have formed during the cosmological evolution of our Universe through the Kibble--Zurek  mechanism\,\cite{Kibble:1976sj,Zeldovich:1974uw,Zurek:1985qw}.
Defects arise as classical non-trivial solutions to the field's equations of motion and physically correspond to extended configurations in space-time. Their formation is naturally expected in Beyond Standard Model (BSM) theories with spontaneously broken gauge or global symmetries. For a group $G$ that breaks to a subgroup $H$, topological defects emerge when the vacuum manifold $\mathcal{M}=G/H$
has a non-trivial homotopy group. In particular, strings form when $\pi_1(\mathcal{M})\neq 1$, while monopoles correspond to $\pi_2(\mathcal{M})\neq 1$.
Topological defects can be a powerful source of gravitational waves (see Ref.\,\cite{Vilenkin:2000jqa} for a review), and are one of the main cosmological targets of current and future detectors, see \emph{e.g.}\,\cite{NANOGrav:2023hvm,LIGOScientific:2025kry,Blanco-Pillado:2024aca,ET:2025xjr}.  

While topological defects are protected by an infinite energy barrier and are thus absolutely stable, non-topological defects may also form but are not guaranteed to remain. 
Famous examples of the latter are semi-local and electroweak strings\,\cite{Vachaspati:1991dz, Vachaspati:1992jk}, whose possible stability relies only on dynamical arguments.
In this paper, we focus on metastable cosmic strings arising from the symmetry-breaking pattern $\mathcal{G} \to G \to H$ with $\pi_2(\mathcal{G}/G) \neq 1$, $\pi_1(G/H) \neq 1$, but $\pi_1(\mathcal{G}/H)= 1$
(see also \,\cite{Preskill:1992ck} for a general classification). Examples where this pattern is realized can be found in the context of Grand Unified Theories\,\cite{Dunsky:2021tih,Buchmuller:2019gfy,Buchmuller:2023aus,King:2020hyd,King:2021gmj,Fu:2023mdu,Buchmuller:2024zzk,Ingoldby:2025wcl}, or describe a pattern related to flavor symmetry breaking\,\cite{Antusch:2025xrs}.
For other recent realizations see 
\cite{Chitose:2024pmz,Chitose:2025cmt}. 
The simplest setup with the required properties is however $\mathcal{G}= SU(2)$, $G= U(1)$, and $H=1$, which is often considered for explicit calculations\,\cite{Preskill:1992ck,Shifman:2002yi}.
As the $G\rightarrow H$ breaking is embedded in a larger group $\mathcal{G} \rightarrow H$ with trivial first homotopy group, the strings forming in the second step are not topologically stable and can decay via the nucleation of a monopole-antimonopole pair. Assuming that the population of monopoles formed in the first step is sufficiently diluted, \emph{e.g.} by cosmic inflation, the life time of the string network is controlled by the ratio of the monopole mass to the string tension, $\kappa \equiv m^2/\mu$.

If the strings are sufficiently long lived, they can have important phenomenological implications. In fact, the gravitational-wave emission from a metastable cosmic-string network has been shown to possibly explain the signal observed at Pulsar Timing Arrays (PTAs)\,\cite{NANOGrav:2023hvm} for a moderate value of $\sqrt{\kappa} \simeq 8$,
which was obtained within the thin-wall approximation for the quantum tunneling. Corrections to the thin-wall limit as studied in\,\cite{Chitose:2023dam}, as well as the fact that 
metastable strings are not infinitely long at the time of formation as recently observed in \cite{Tranchedone:2026lav}, can however shift the best-fit value of $\kappa$, \emph{e.g.} to $\sqrt{\kappa} \gtrsim 30$ according to\,\cite{Tranchedone:2026lav}
(see \cite{Asl:2026zpj} for a very recent work revisiting this issue and the related gravitational wave spectrum).

When discussing metastable strings, one implicitly assumes that the strings are actually stable at the classical level, or, in other words, that small fluctuations around the string background have a positive mass squared and can not grow. Interestingly, the classical stability of the string solutions in the simplest model with $SU(2) \rightarrow U(1) \rightarrow 1$ introduced in\,\cite{Shifman:2002yi}
has not yet been thoroughly investigated.
Providing such analysis is the main goal of this paper. In particular, one can wonder what is the needed hierarchy between the two symmetry breaking scales that guarantees the absence of unstable modes in the string solution, which is of the Abrikosov-Nielsen-Olesen (ANO) type\,\cite{Nielsen:1973cs, Abrikosov:1956sx} in this model, and how this correlates with the monopole mass and string tension.

To answer this question, we will study in detail the linearized equations for the infinitesimal
perturbations around the metastable string solution, and find under which conditions they do not contain any negative eigenvalue.
In doing so, we will employ similar methods to the ones extensively used in the study of 
electroweak strings in the SM and its extensions\,\cite{Achucarro:1999it, Goodband:1995he, Earnshaw:1993yu}
(see also e.g. \cite{Hindmarsh:1991jq,James:1992wb,James:1992zp,Achucarro:1993bu, Garaud:2010ng,
Masperi:1993fw,Eto:2016mqc,Abe:2020ure,Eto:2021dca,
Eto:2024xvc,Bian:2026tco, Forgacs:2016dby,Forgacs:2019tbn, Kanda:2023yyz}).

Focusing on the model parameter space where the PTA signal can be explained by metastable strings according to the analysis of \cite{NANOGrav:2023hvm}, we find regions where the strings are indeed classically stable and this standard analysis applies unchanged. However, we also identify portions of the parameter space where the strings are classically unstable.
This remains true also when considering a shorter string network life time  as the cases discussed in\,\cite{Tranchedone:2026lav,Asl:2026zpj}.

What actually happens in the regions of classical instability that would otherwise lead to a viable explanation of the PTA signal remains an open question. In fact, the strings may just dissolve soon after the network formation, or survive rearranging into a different type of string profile.
The evolution of the string following the instability can not be addressed at the level of fluctuations, and thus goes beyond the scope of this paper. Nevertheless, we will provide a study of this phenomenon within a simplified version of the model. Our results hint to the survival of the network, with the strings approaching a new profile with a scalar condensate and a smaller tension.
In this scenario, the strings could still explain the PTA signal, even though their tension and decay rate would need to be reassessed in order to make a precise statement.

\medskip

The rest of the paper is structured as follows. In section \ref{Section Metastabel Strings and GWs} we review the basics of metastable strings and their GW spectrum, and how it can fit the PTA signal.
In section \ref{Section Model} we introduce the model and define the fundamental parameters dictating the microscopic properties of the string. We follow this discussion with a brief presentation of the monopole and string solutions together with their energetics in section \ref{Section topological defects}. In section \ref{Section stability} we perform a linear stability analysis about the ANO string and identify which field fluctuations produce unstable modes. This is followed in section \ref{Section numerical results} by a discussion of our numerical results from which we identify the different instabilities that are present and explain their physical origin. We then also translate our results to predict their phenomenological implications, and in particular for which values of the fundamental parameters the metastable strings may no longer fit the PTA data. 
Finally, we present a simplified analysis of the string profile evolution following the classical instability in section \ref{Section chasing the instability}, and conclude in section \,\ref{sec:conclusion}.

\section{Metastable Strings and Gravitational Waves}\label{Section Metastabel Strings and GWs}

In this section we give a short review on the gravitational wave background generated by metastable strings and its detectability at current and future experiments.
In particular, metastable strings have been identified as one of the most promising candidates for accommodating the signal observed at PTAs \cite{NANOGrav:2023hvm}. 

\begin{figure}
    \centering
    \includegraphics[width=7.32cm]{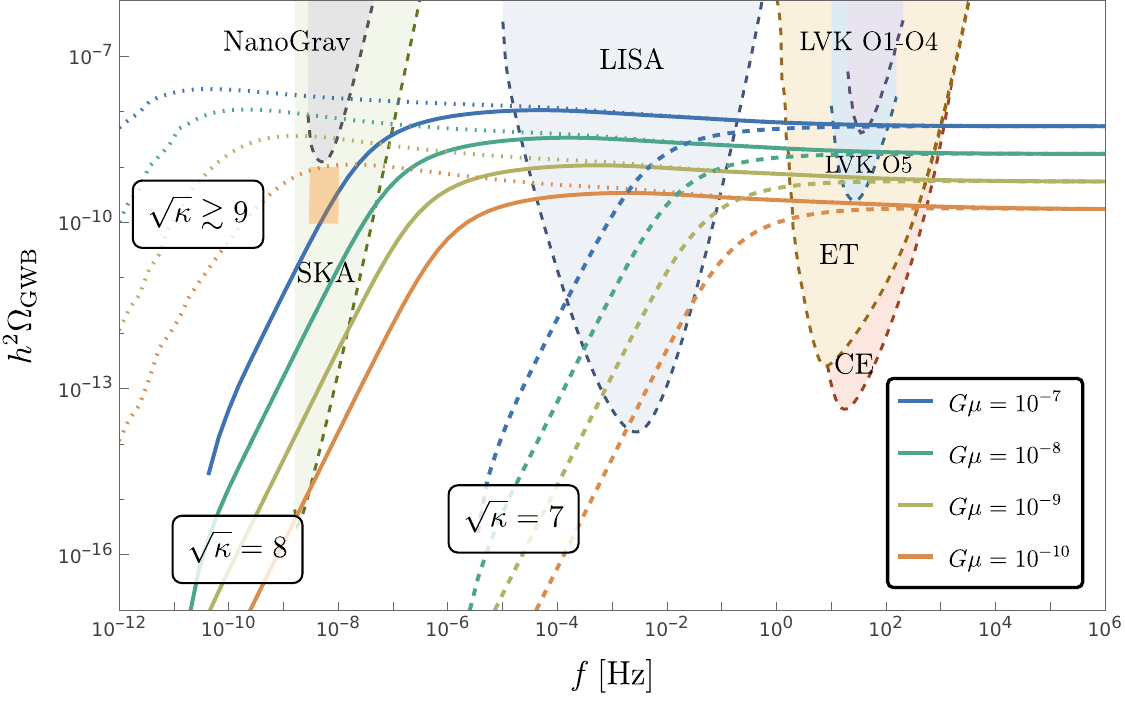}
    ~
      \includegraphics[width=7.45cm]{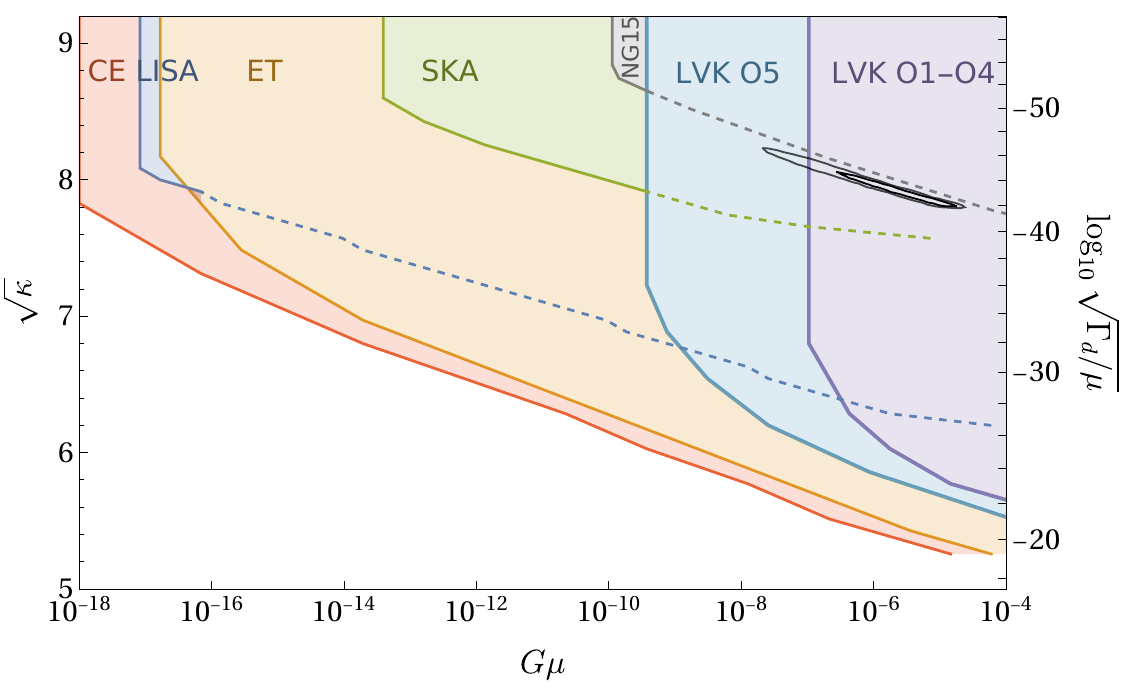}
    \caption{\textbf{Left:} The GW spectrum from metastable cosmic strings with loop size $\alpha=0.1$ for different values of $G\mu$ and $\sqrt{\kappa}$. The solid lines correspond to $\sqrt{\kappa}=8$, dashed lines to $\sqrt{\kappa}=7$ and the dotted lines to those from stable strings whose decay time is larger than the age of the Universe. In the background, the $\text{SNR}=1$ sensitivity curves for LVK run O1-O4a and for run O5 have been plotted \cite{LIGOScientific:2025bgj}, as well as for ET \cite{ET:2025xjr}. In addition, the sensitivities of LISA, CE and SKA from \cite{Schmitz:2020syl} are also shown. \textbf{Right:} The corresponding parameter scan in $G\mu$ and $\sqrt{\kappa}$. The overlapping regions represent region in the $G\mu$ and $\sqrt{\kappa}$ space for which the spectrum of the string network can be detected for the different GW experiments just mentioned. The black region denotes the preferred region of the 15yr observation time of NANOGrav, taken from \cite{NANOGrav:2023hvm}.
    }
    \label{Fig GW spectrum from metastable strings}
\end{figure}

A network of metastable strings 
emits gravitational waves through the continuous formation and dissipation of loops, as standard gauged cosmic strings (see e.g. \cite{Gouttenoire:2019kij} for a review and references therein). 
However, due to the non-topological nature of the metastable strings, they can decay via spontaneous nucleation of monopole anti-monopole pairs. Therefore, depending on the microscopic properties (monopole mass and string tension), the finite life time of the metastable string network implies an exponential suppression of the loop number density at the time of decay, resulting in an infrared cutoff in the GW spectrum. Such modified spectrum was derived in \cite{Buchmuller:2021mbb, Buchmuller:2020lbh} (see also \cite{Buchmuller:2023aus,Leblond:2009fq}), and we outline the main ingredients in Appendix \ref{app:reviewGWmetastable}. 
The relevant quantity controlling the spectrum is the time of network decay. 
Assuming infinitely long strings and considering the thin wall approximation, the time 
of decay 
is determined by the decay rate per unit length\,\cite{Preskill:1992ck,Leblond:2009fq}:
\begin{equation}
\label{eq:thin_wall}
    \Gamma_d = \frac{\mu}{2\pi}\text{e}^{-\pi\kappa}, \qquad \kappa \equiv \frac{m^2}{\mu},
\end{equation}
where $m$ is the mass of the monopole, and $\mu$ the tension of the string. Assuming radiation domination and considering Horizon-size strings, one can convert this into a decay temperature:
\begin{equation}
T_d \simeq 0.4 \text{~MeV} 
\left(\frac{G \mu}{10^{-7}} \right)^{1/4} \left(
\frac{\sqrt{\Gamma_d / \mu}}{10^{-40}}
\right)^{1/2}
\left( \frac{g_{\star}}{10}
\right)^{-1/4}.
\end{equation}
Representative GW spectra for different values of $\sqrt{\kappa}$ and of the string tension are reported in Figure \ref{Fig GW spectrum from metastable strings} (left). On the same figure, we also show the Power Law Integrated (PLI) sensitivity curves of current and future gravitational wave experiments, respectively 
LIGO-Virgo-KAGRA\footnote{For the most recent LVK bounds on stable cosmic strings see \cite{LIGOScientific:2025kry}.}
(curve taken from \cite{LIGOScientific:2025bgj}), ET \cite{ET:2025xjr}, CE and LISA from \cite{Schmitz:2020syl}. For PTAs we show in the same figure the region favored by the NANOGrav collaboration \cite{NANOGrav:2023hvm} for the detected gravitational wave background.
The reach of GW experiments for the metastable string network is further elucidated on the right panel of Figure \ref{Fig GW spectrum from metastable strings}.

For a ratio of monopole mass over string tension of order $\sqrt{\kappa} \simeq 8$ the GW signal from metastable strings can explain the PTA data. 
For large values of $\sqrt{\kappa}$ the contours become vertical since the lifetime of the strings is so large that the IR cutoff is outside the sensitivity range of the corresponding experiment, and in such case the limit on $G \mu$ corresponds to the one for stable cosmic strings.
For small values of $\sqrt{\kappa}$ the lifetime is so short that the signal is suppressed and no detection is possible.
On the same Figure, we display the one and two sigma contours as derived by the NANOGrav analysis \cite{NANOGrav:2023hvm}. The favorite region is only partially excluded from LVK, and possibly at reach of the next LVK run.

We emphasize that the functional relation between 
the decay temperature and the monopole over string tension 
could receive important corrections with respect to the simple expressions in \eqref{eq:thin_wall}, eventually modifying the preferred value of $\sqrt{\kappa}$ for PTAs.
In addition to possible corrections beyond the thin wall regime\,\cite{Chitose:2023dam}, it has been recently pointed out  
that the strings are already broken into super-horizon segments at formation \cite{Tranchedone:2026lav},
effectively shortening the lifetime of the network and moving the best-fit $\sqrt{\kappa}$ to larger values.
Additionally, it might be possible that monopoles from the first step of the symmetry breaking re-enter the horizon at later stages, effectively destroying the strings network before their proper quantum decay time \cite{Lazarides:2022jgr,Tranchedone:2026lav}.
Our classical stability analysis in the next sections will therefore cover a sufficiently large range of $\kappa$ to include these cases on top of the standard predictions.

\section{The model realizing $SU(2)\to U(1) \to 1$}
\label{Section Model}

The $SU(2)$ gauge theory containing metastable cosmic strings that we consider was first proposed in \cite{Preskill:1992ck}, and later refined in \cite{Shifman:2002yi}. The model leads to a sequence of two phase transitions. In the first step, the $SU(2)$ symmetry breaks down to a $U(1)$ subgroup. The latter then breaks down leaving no residual symmetry behind. Schematically, if we associate two breaking scales $V$ and $v$ to the phase transitions, we can represent the chain of spontaneous symmetry breaking by
\begin{equation}
        SU(2) \xrightarrow[]{\: V\:} U(1) \xrightarrow[]{\: v\:} 1.
        \label{Schematic breaking of the model SU(2) to U(1) to 1}
\end{equation}
Such a model may be the low-energy effective field theory of a Grand Unified Theory, or it may as well correspond to a completely dark sector disconnected from the Standard Model. We will not make any assumptions on the specifics, and shall keep our discussion general. In what follows we adopt the mostly minus metric convention
$(g_{\mu\nu}) = (1,-1,-1,-1)$.

The Lagrangian of the theory is given by:
\begin{equation}
    \mathcal{L} = \frac{1}{2}(D_\mu\phi^{a})(D^{\mu}\phi^{a}) + (D_{\mu}h)^{\dagger}(D^{\mu}h) - \frac{1}{4}F_{\mu\nu}^{a}F^{a\,\mu\nu} - V(\phi, h),
    \label{Lagrangian full theory OWN model}
\end{equation}
where $\phi^a$ ($a=1,2,3$) is a real scalar triplet in the adjoint representation of $SU(2)$, and $h=(h_1, h_2)^T$ is a complex scalar doublet. The covariant derivatives are therefore
\begin{align}
    D_\mu h &= \partial_{\mu}h - \frac{ig}{2} \, \sigma^{a} A_{\mu}^{a}h,\\
    D_{\mu}\phi^{a} &= \partial_{\mu}\phi^{a} + g \,\epsilon^{abc}A_{\mu}^{b}\phi^{c},
\end{align}
where $g$ is the gauge coupling constant. The non-abelian field strengths are given by the usual anti-symmetric combination in the gauge fields:
\begin{equation}
    F_{\mu\nu}^{a} = \partial_{\mu}A_{\nu}^{a} - \partial_{\nu}A_{\mu}^{a} + g\, \epsilon^{abc}A_{\mu}^{b}A_{\nu}^{c}.
\end{equation}
Finally, other than including two Mexican hat-like potentials for the two scalar sectors, $V(\phi, h)$ also contains a portal term connecting the sectors together and ensuring they are charged under the same $SU(2)$ gauge group. Explicitly, the potential reads
\begin{equation}
    V(\phi, h) = \lambda \left(h_1^{\dagger}h_1 + h_2^{\dagger}h_2 - v^2\right)^2 + \tilde{\lambda} \left(\phi^{a}\phi^{a} - V^{2}\right)^{2} + \gamma \left|\left(\frac{\phi^{a}\sigma^{a}}{2} - \frac{V}{2}\right)h\right|^2,
    \label{Potential full theory own model}
\end{equation}
where $\lambda, \tilde{\lambda}$ are the self-interaction couplings for the two sectors, and $\gamma$ dictates the interaction strength between the adjoint and fundamental scalars as modeled by the portal. 

We shall not analyze in detail the cosmological history of the model, which might depend also on the scalar couplings with other particles in the thermal bath.
One possibility is that,
as the Universe expands and cools down, a first transition occurs around a temperature $T\sim V$ leading to the symmetry breaking $SU(2)\to U(1)$, and subsequently a second transition occurs at $T \sim v$ where the remaining $U(1)$ is broken. An intermediate period of inflation is typically required to dilute the population of monopoles formed in the first step. 
We notice however that the results of our analysis do not depend on the precise mechanism that dilutes the monopole population.

In the first breaking step,
we align without loss of generality the scalar triplet along the $\phi^3$ direction such that $\phi^3$ acquires a mass $m_{\phi^3}$ given by
\begin{equation}
    \braket{\phi^a} = V \delta^{a,3} \quad \text{with} \quad m_{\phi^3} = \sqrt{8\tilde{\lambda}}\, V.
\end{equation}
Consequently, the remaining scalars $\phi^{1}$ and $\phi^2$ become massless would-be Goldstone bosons around the new vacuum which are absorbed as an extra longitudinal polarization degree of freedom according to the Brout-Englert-Higgs mechanism by the gauge fields $A^{1}$ and $A^2$. This leads to the generation of a mass term for these gauge fields,
\begin{equation}
m_{W}\equiv m_{A^{\pm}} = gV,
\end{equation}
and we defined the complex combinations $A^{\pm}_\mu$ and $\phi^{\pm}$ as
\begin{equation}
    A^{\pm}_\mu = \frac{1}{\sqrt{2}}\left(A_\mu^{1} \mp i A_{\mu}^{2} \right) \quad \text{and} \quad \phi^{\pm} = \frac{1}{\sqrt{2}}\left(\phi^{1} \mp i \phi^{2} \right)
\end{equation}
We also label the mass of $A^\pm$ field as a $W$-boson mass for analogies with the electroweak sector of the Standard Model.
For energy scales below $m_W$ and $m_{\phi^3}$, we are left with a $U(1)$ gauge symmetry.
Moreover, the vacuum expectation value of $\phi^3$ induces through the portal term a mass for the second component of the doublet field,
\begin{equation}
    m_{h_2} = \sqrt{\gamma} V.
\end{equation}
When the Universe reaches $T\sim v$, the remaining $U(1)$ symmetry breaks spontaneously. The vacuum of the fundamental scalars is then given by
\begin{equation}
    \braket{h} = \begin{pmatrix}
        v \\ 0
    \end{pmatrix} .
\end{equation}
This final breaking leads to the following masses,
\begin{equation}
    m_{h_1} = 2\sqrt{\lambda} v, 
    \quad m_Z \equiv m_{A_3} = gv/\sqrt{2},
\end{equation}
where we label the $A^3$ field as a $Z$-boson for analogies with 
the 
$Z$-string of the electroweak sector \cite{Vachaspati:1992jk}.

In view of the later computations, we define the following set of four fundamental dimensionless parameters through the ratio of the relevant mass scales and couplings in the problem:
\begin{align}
    \beta_M &= \left(\frac{m_{\phi^3}}{m_W}\right)^2 = \frac{8\tilde{\lambda}}{g^2} && \alpha = \left(\frac{m_{h_2}}{m_Z}\right)^2 = \frac{2\gamma V^2}{(gv)^2} \nonumber \\
    \beta_S &= \left(\frac{m_{h_1}}{m_Z}\right)^2 = \frac{8\lambda}{g^2}  && \eta = \left(\frac{m_W}{m_Z}\right)^2 = \frac{2 V^2}{v^2}
    \label{Fundamental parameters definition}
\end{align}

\noindent
Having introduced the model, in the following section, we present the two essential ingredients that arise from the two phase transitions: the monopoles made up from the adjoint scalar sector and the cosmic strings from the fundamental sector. Of particular interest, are the string solutions that can be constructed along the field $h_1$ emerging from the breaking $U(1)\to 1$. In later sections, we study the classical stability of the string solution embedded in the complete theory in function of the fundamental parameters of the model listed above. 

Although the stability of the string solution is not topologically protected, we will be able to find regions of the parameter space where the string  is stable even without having a large hierarchy between the $SU(2)$ breaking scale and the $U(1)$ breaking scale.
In order to be consistent with the assumed cosmological history, we will nevertheless focus on values of $\eta \gtrsim 1$.

\section{The topological defects}\label{Section topological defects}

The model described in section \ref{Section Model} leads to the formation of defects.
The first breaking step $SU(2)\to U(1)$ leads to a vacuum manifold $\mathcal{M}_1=SU(2)/U(1)$, allowing for the formation of monopoles. The second symmetry breaking $U(1) \to 1$ has a vacuum manifold $\mathcal{M}_2 = U(1)$, such that $\pi_1(\mathcal{M}_2)=\mathbb{Z}$. This last breaking results in the formation of a cosmic string network in the early Universe. We stress again that the
stability of these cosmic strings is not guaranteed by topological arguments given that the net vacuum manifold $\mathcal{M}=SU(2)$ has a trivial first homotopy group $\pi_1 (SU(2))=1$
and hence one expects that the string solution can dissipate either classically or quantum mechanically by exploring the full scalar manifold.
Before addressing the stability of the string solutions, in this section, we review the profiles and the energy properties of the monopole (its mass) and of the string (its tension) emerging in this model.

\subsection{Monopole profile and mass function}\label{Appendix Monopoles}
\begin{figure}
    \centering
    \includegraphics[width=7.3cm]{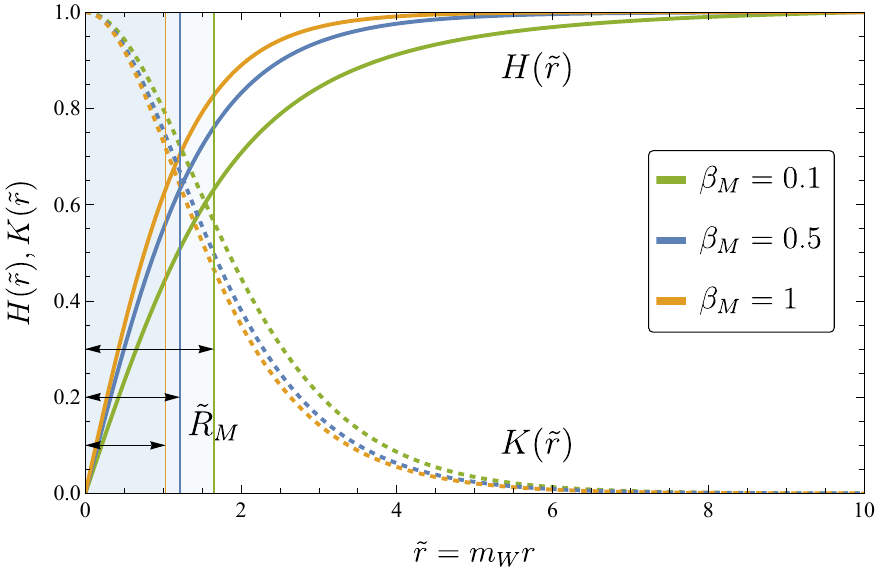}
    ~~
    \includegraphics[width=7.3cm]{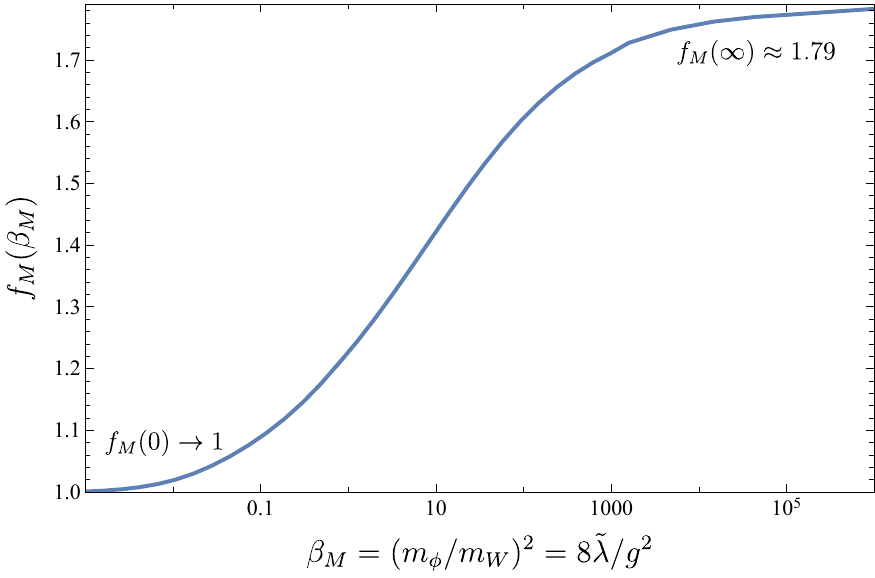}  
    \caption{\textbf{In the left panel:} The profile solutions $H(\tilde r)$ and $K(\tilde r)$ of Eq. \eqref{Eq ansatz monopole} for different values of $\beta_M = (m_{\phi^3}/m_W)^2$. The value $\tilde R_M$ is defined as the value for which $H(\tilde{R}_M) \equiv 1-e^{-1}$, and indicates a value for the monopole radius.
    \textbf{In the right panel:} the monopole mass function integrated via Eq. \eqref{Numerical function monopole mass} as a function of the parameter $\beta_M$.}
    \label{Fig Monopole profiles and mass}
\end{figure}

We now review the explicit form, profiles and mass function of the monopoles formed during the first phase transition $SU(2)\to U(1)$. For this purpose, we omit the presence of the fundamental scalars and only focus on the dynamics of the adjoint scalars. Using the same notation as in Section \ref{Section Model} the theory reduces to
\begin{equation}
    \mathcal{L} \supset \frac{1}{2}(D_{\mu}\phi^{a})(D^{\mu}\phi^{a}) - \frac{1}{4}F^{a}_{\mu\nu}F^{a\mu\nu} - \tilde{\lambda}\left(\phi^{a}\phi^{a} - V^2\right)^{2},
    \label{eq:partformonopole}
\end{equation}
with as equations of motion:
\begin{align}
    D_{\mu}D^{\mu}\phi^{a} &= -4\tilde{\lambda}\left(\phi^{b}\phi^{b} - V^2\right) \phi^{a}, \label{Eq phi monopole} \\
    D_{\nu}F^{a\,\mu\nu} &= g\epsilon^{abc}\phi^{b} D^{\mu}\phi^{c}, \label{Eq A Monopole}
\end{align}
Initially proposed by 't Hooft and Polyakov in respectively \cite{tHooft:1974kcl} and \cite{Polyakov:1974ek}, the monopole solution arises by considering a static and spherically symmetric ansatz for the scalar fields. In particular, the monopole ansatz mixes the field space indices $a$ with the coordinate indices $i$, effectively resulting in the so-called hedgehog configuration. The 't Hooft-Polyakov monopole ansatz is given by \cite{Shifman:2012zz}
\begin{equation}
    \phi^{a} = V H(r) \frac{x^{a}}{r}, \:\:\:\:\: A_{0}^{a} = 0 
    \:\:\:\:\: \text{and} \:\:\:\:\: A_{i}^{a} = \frac{\epsilon^{aij} x^{j}}{g \,r^{2}} (1-K(r)), 
    \label{Eq ansatz monopole}
\end{equation}
where $r^2 = x_1^2 + x_2^2 +x_3^2$ is the spherical radius of the space coordinates\footnote{Note that the monopole ansatz may differ from e.g. \cite{Kirkman:1981ck, Shnir:2005vvi} where the profile for $\phi^a$ is defined with an extra $r$ factor in the denominator: $H(r) \to H(r)/r$.}. The exact form of the gauge fields is chosen in such a way that it cancels any diverging energy at infinity, thus evading Derrick's theorem \cite{Derrick:1964ww, Vilenkin:2000jqa} of soliton configurations of infinite energy. Therefore, the boundary conditions on the profiles $H(r)$ and $K(r)$ must satisfy:
\begin{equation}
    H(0) = 0,\quad  K(0) = 1, \quad H(\infty)\to1, \quad K(\infty)\to0
\end{equation}
Substituting the monopole ansatz in the equations of motions listed in Eqs \eqref{Eq phi monopole} and \eqref{Eq A Monopole}, and rescaling the radius according to $r \to \tilde{r} \equiv  m_W r = gVr$, we obtain the following equations for the profiles:
\begin{align}
    \frac{d^{2}H}{d\tilde{r}^2} + \frac{2}{\tilde{r}}\frac{dH}{d\tilde{r}} = \frac{2}{\tilde{r}^2} HK^{2} -
    \frac{\beta_{M}}{2}\left(1-H^2\right)H, \\
    \frac{d^{2}K}{d\tilde{r}^{2}} = \frac{1}{\tilde{r}^{2}}(K^{2} - 1)K + K H^{2}.
    \label{EoMs monopole profile}
\end{align}
The profiles are computed via a relaxation algorithm for different values of the ratio $\beta_M$ defined in Eq. \eqref{Fundamental parameters definition}. The results for several values of $\beta_M$ are shown in the left panel of Figure \ref{Fig Monopole profiles and mass}.

\paragraph{Monopole size} While the monopole size $R_M$ scales with $R_M\sim m_{\phi^3}^{-1}$, an exact definition of this distance is a mere convention. For the profiles, we define the monopole size $R_M$ to be the distance for which in rescaled units $H(\tilde{R}_M) \equiv 1-e^{-1}$. These values are illustrated in Figure \ref{Fig Monopole profiles and mass} together with the different profiles.

\paragraph{Monopole mass function} Finally, the mass of the monopole can be computed by integrating over the total energy density (i.e. the Hamiltonian density) of the theory. Substituting the monopole ansatz followed by the rescaling used earlier we find that the monopole mass equals
\begin{equation}
    m = \frac{4\pi V}{g} f_M(\beta_M),
\end{equation}
where the function $f_M(\beta_M)$ is a  dimensionless function defined as the integral:
\begin{equation}
    f_{M}(\beta_M) = \int \tilde{r}^{2}d\tilde{r} \left[\frac{1}{2}\left( \frac{dH}{d\tilde{r}}\right)^2 + \frac{H^2 K^2}{\tilde{r}^2}  + \frac{\left(1-K^2\right)^2}{2\tilde{r}^4}+\frac{1}{\tilde{r}^2}\left(\frac{dK}{d\tilde{r}}\right)^2 + \frac{\beta_{M}}{8}\left(H^2 - 1\right)^2 \right].
    \nonumber\label{Numerical function monopole mass}
\end{equation}
The integral is obtained by a standard numerical integration using the previously obtained profiles. The result of the integration for different values of $\beta_M$ is shown in the right panel of Figure \ref{Fig Monopole profiles and mass}, and agrees with the monopole mass obtained in \cite{Chitose:2023dam}. In particular, the mass function $f_M(\beta_M = 0) = 1$ and it approaches $f_M \approx 1.79$ as $\beta_M \to \infty$, also in agreement with e.g. \cite{Shifman:2012zz, Chitose:2023dam, Kirkman:1981ck}.

\subsection{String solution and its tension}\label{Section String Profile}
We now review the string solutions that arise in the low-energy theory of the model, where the remaining $U(1)$ gauge symmetry is spontaneously broken.
In this regime, the non-abelian part of the gauge field is expected to be frozen and the adjoint scalar sector fixed to its vacuum expectation value, $\phi^{a} = V \delta^{a3}$. As such, in this low-energy effective theory, only the $h_1, h_2$ and $A_\mu^3$ fields are dynamical\,\footnote{Notice that we keep the field $h_2$ in the low-energy theory. In fact, this field can be light even for large $V$ when allowing for a small portal coupling $\gamma$.}. The low-energy effective theory is then described by the Lagrangian
\begin{equation}
\label{eq:partforstring}
    \mathcal{L}_{\text{eff}} \supset (D_{\mu}h_1)^{\dagger} D^{\mu}h_1 + (D_{\mu}h_2)^{\dagger} D^{\mu}h_2 -\frac{1}{4}F_{\mu\nu}^{3} F^{3\,\mu\nu} - V_{\text{eff}}(h_{1}, h_{2}).
\end{equation}
This effective description corresponds to an abelian gauge theory for which the field strength is the same as that for electrodynamics, but in the $A^3_\mu$ component:
\begin{equation}
    F_{\mu\nu}^3 = \partial_{\mu}A_{\nu}^{3} - \partial_{\nu}A_{\mu}^3.
\end{equation}
The covariant derivatives now act on the fundamental scalars as
\begin{align}
     D_{\mu}h_{1} &= (\partial_{\mu} - \tfrac{i}{2}gA_{\mu}^{3})h_{1}, \\
     D_{\mu}h_{2} &= (\partial_{\mu} + \tfrac{i}{2}gA_{\mu}^{3})h_{2},
\end{align}
Lastly, the effective potential becomes a single Mexican hat potential for the fundamental scalar sector and contains an extra effective mass term for $h_2$;
\begin{equation}
\label{eq:effectivePot}
    V_{\text{eff}}(h_{1}, h_{2}) = \lambda\left(|h_1|^2 + |h_{2}|^{2} - v^{2}\right)^{2} + \gamma V^{2} |h_{2}|^{2}.
\end{equation}
Taking into account that the vacuum of the theory enforces $\braket{h}=(v,0)^T$, the string solution is constructed along the $h_1$ field.
Following the lines of \cite{Abrikosov:1956sx, Nielsen:1973cs, Vilenkin:2000jqa}, the cosmic string solution is obtained by considering the following Abrikosov-Nielsen-Oleson (ANO) static, cylindrically symmetric string ansatz in the fields:
\begin{equation}
    h_1 = vf(\rho) e^{in\theta}, \quad \quad h_2 = 0, \quad A^{a}_\mu = \frac{2n}{g\rho}\zeta(\rho)\, \delta^{a3}\,\delta_{\mu\theta}, 
    \label{Standard string ansatz}
\end{equation}
where $\rho^2 = x_1^2 + x_2^2$ is the cylindrical distance such that the string is aligned with the $3$-axis. The gauge field has been chosen in such a way that the energy contribution from the covariant derivatives vanishes at infinity: $D_\mu h_1 \to 0$ as $\rho\to\infty$. The functions $f(\rho)$ and $\zeta(\rho)$ are the scalar and gauge field profiles respectively, which need to be determined numerically through the equations of motion. For regularity in the origin and to ensure a finite energy configuration at infinity, the profiles must satisfy the boundary conditions:
\begin{equation}
    f(0) = \zeta(0) = 0, \quad f(\infty),\:\: \zeta(\infty)\to 1. 
\end{equation}
Finally, $n$ is the winding number defining the vortex geometry. Plugging this ansatz into the equations of motion, and rescaling the radial distance according to 
\begin{equation}
    \rho \to \tilde{\rho} \equiv m_Z\rho  = (g v/\sqrt{2})\rho,
    \label{Rescale of coordinate in Low E theory}
\end{equation}
yields the following two equations in the profiles $f(\tilde{\rho})$ and $\zeta(\tilde{\rho})$:
\begin{align}
    f''(\tilde{\rho}) + \frac{1}{\tilde{\rho}}f'(\tilde{\rho}) - \frac{n^2}{\tilde{\rho}^2} \left(1-\zeta(\tilde{\rho})\right)^2f(\tilde{\rho}) - \frac{\beta_S}{2} \left(f(\tilde{\rho})^2 - 1 \right)f(\tilde{\rho}) &= 0,\\
   \zeta''(\tilde{\rho}) - \frac{1}{\tilde{\rho}}\zeta'(\tilde{\rho}) + f^2(\tilde{\rho})\left(1-\zeta(\tilde{\rho})\right) = 0.
   \label{Eqns of local string for f and zeta}
\end{align}
A full solutions to these equations needs a numerical implementation, which was done for different values of $\beta$ via relaxation.

\begin{figure}
    \centering
    \includegraphics[width=7.3cm]{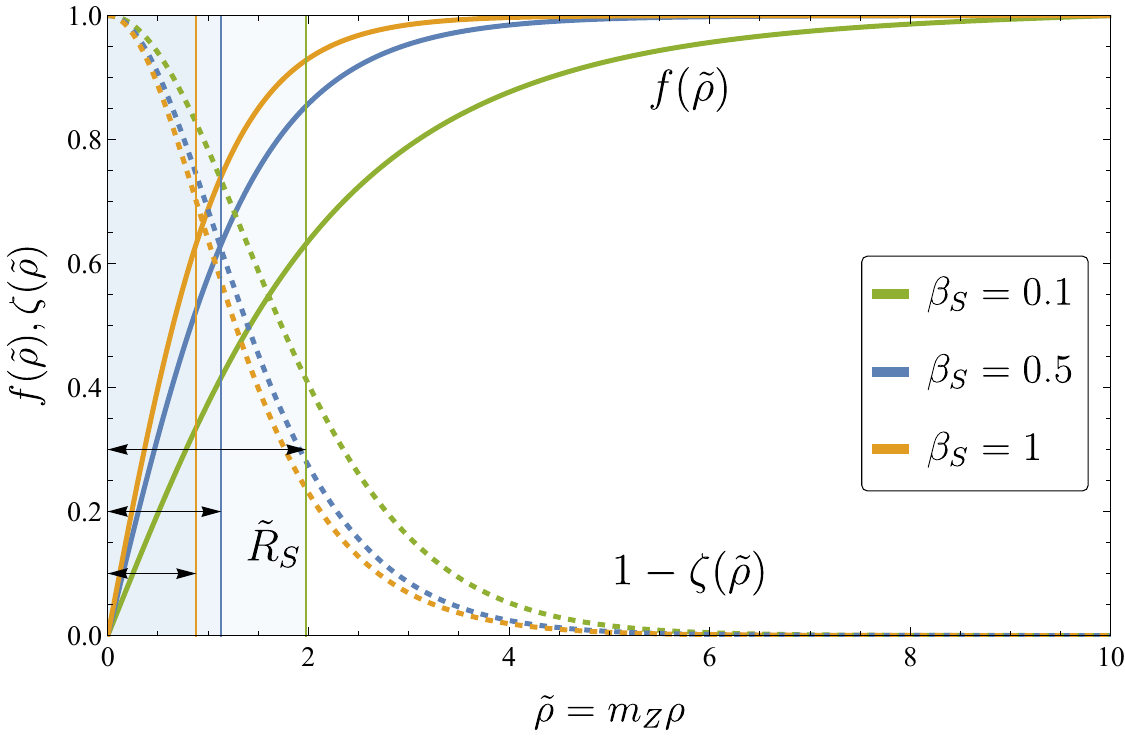}
    \includegraphics[width=7.3cm]{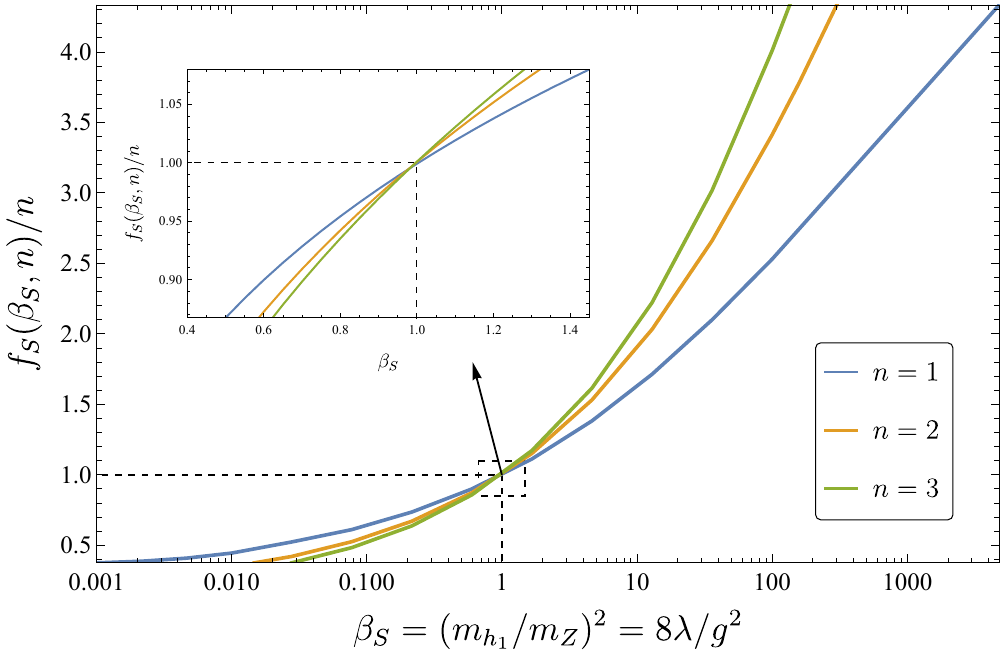}
    \caption{\textbf{In the left panel:} The profile solutions $f(\tilde \rho)$ and $\zeta(\tilde \rho)$ of Eq. \eqref{Standard string ansatz} for different values of $\beta_S = (m_{h_1}/m_Z)^2$ and for unit winding $n=1$. The value $\tilde R_S$ is defined as the value for which $f(\tilde{R}_S) \equiv 1-e^{-1}$, and indicates a value for the string width. \textbf{On the right:} the string tension function integrated via Eq. \eqref{Numerical function string tension} as a function of the parameter $\beta_S$.}
    \label{Fig String Profiles and tension}
\end{figure}

\paragraph{String width} Similarly to the  monopole, the width of the string $R_S$ scales with $R_S\sim \max(m_{h_1}^{-1}, m_{Z}^{-1})$, an exact definition of which is a convention. For our profiles, we define the string width $R_S$ to be the distance for which in rescaled units $f(\tilde{R}_S) \equiv 1-e^{-1}$. These values are illustrated together with their profiles in the left panel of Figure \ref{Fig String Profiles and tension}.

\paragraph{String tension}
The string tension $\mu$ (i.e. the energy per unit length) is obtained by integrating over the Hamiltonian density associated to the low-energy theory in \eqref{eq:partforstring}. Substituting the string ansatz and rescaling according to Eq. \eqref{Rescale of coordinate in Low E theory}, we find that:
\begin{equation}
    \mu = 2\pi v^2 f_S(\beta_S,n),
    \label{String tension}
\end{equation}
where the function $f_S$ is a function of the dimensionless parameter $\beta_S$ defined in Eq. \eqref{Fundamental parameters definition} and $n$ the winding number. Explicitly, the function is found to be the following integral:
\begin{equation}
    f_S(\beta_S,n) = \int_{0}^{\infty}\tilde\rho\,d\tilde\rho \left[ \left(\frac{df}{d\tilde\rho}\right)^{2} + \frac{n^2}{\tilde\rho^2}\left(\zeta-1\right)^2 f^2 + \frac{n^2}{\tilde\rho^2}\left(\frac{d\zeta}{d\tilde\rho}\right)^2 + \frac{\beta_S}{4}\left(f^2 - 1\right)^2  \right] ,
    \label{Numerical function string tension} \nonumber
\end{equation}
which we compute by a standard numerical integration scheme using the profiles obtained earlier. The result of this function for different values of $\beta_S$ and of the winding number is shown in the left panel of Figure \ref{Fig String Profiles and tension}. We observe that for $\beta_S$ of order unity the mass function behaves like $f_S(\beta_S\approx 1, n) \approx n$, (as can also be seen in the zoomed in portion of Figure \ref{Fig String Profiles and tension}). At last, for small values of $\beta_S$, the dimensionless string tension function behaves like 
\cite{Shifman:2012zz}
\begin{equation}
    f_S(\beta_S \to 0, 1) \to \frac{1}{\ln\beta_S^{-1}},
\end{equation}
while for large values of $\beta_S$, the string tension diverges logarithmically according to
\begin{equation}
    f_S(\beta_S \to \infty, 1) \to \ln \beta_S.
\end{equation}

\paragraph{Magnetic flux} Lastly, we also compute the magnetic flux associated to the string configuration. The magnetic field associated to $A^3$ is fully aligned on the $z$-axis and is given by $B^3_z = F_{12}^3 = \partial_xA^3_y - \partial_y A^3_x$. The corresponding magnetic flux $\Phi_B$ is then obtained by a surface integration over the magnetic field over the $xy$-plane, which, using Stokes' theorem, becomes a line integral evaluated at spatial infinity:
\begin{equation}
    \Phi_B = \int d^2x \: B_z = \oint_{\rho=\infty} A^{3}_i dx^i = \int_{\theta=0}^{\theta = 2\pi} \left(\frac{2n}{g\rho}\right) \left(\rho d\theta\right) = \frac{4\pi n}{g} 
    \label{Magnetic flux of cosmic string}
\end{equation}
and the flux is thus necessarily quantized. Furthermore, it is conserved in the case where the strings are topological.

\section{Stability analysis}
\label{Section stability}
In this section we investigate the classical stability of the string solution whose ansatz was given in Eq.\,\eqref{Standard string ansatz} with winding number $n=1$. 
When considered in the full model, the strings are non-topological and their stability is no longer guaranteed. Stability can thus only be achieved dynamically, and only a detailed stability analysis around the background string configuration will reveal the distinction between classically stable and unstable strings.
A classical instability would imply that the strings immediately dissolve at the $U(1)$ phase transition (analogously to what occurs for electroweak strings), thereby not sourcing any gravitational wave signal. Alternatively, the instability of the basic ANO profile may result in a new stable configuration, which should then be used as the initial state for the quantum tunneling.
Therefore, the analysis presented in Ref.\,\ref{Section Metastabel Strings and GWs} only applies to strings that are classically stable.

\subsection{Linearized equations of motion}
In order to probe the classical stability of the string ansatz, we consider perturbations of all fields on top of the background string configuration. Therefore, all fields $\Psi=\{h,h^\dagger, \phi^\pm, \phi^3, A_\mu^\pm, A^3_\mu\}$ must be considered as dynamical in our analysis. This will allow the strings to possibly decay via perturbations exploring the non-abelian directions in field space. 
The theory is characterized by the following equations of motion: 
\begin{align}
    D_{\mu}D^{\mu}\phi^{a} &= -4\tilde{\lambda}\left(\phi^{b}\phi^{b}-V^2\right)\phi^{a} - \frac{\gamma}{2}h^{\dagger}\left(\phi^{a}-V\sigma^{a}\right)h \label{EOM Full own model in h},\\
    D_{\mu}D^{\mu}h &= -2\lambda\left(h^\dagger h -v^2\right)h -\frac{\gamma}{4}\left(\phi^a \sigma^a - V\right)^2 h,\\
    D_{\nu}F^{a\mu\nu} &= g\epsilon^{abc}\phi^{b}D^{\mu}\phi^{c} + \frac{ig}{2} \left(h^{\dagger}\sigma^{a}D^{\mu}h - (D^{\mu}h)^{\dagger} \sigma^{a} h \right),\label{EOM Full own model in A}
\end{align}
with $a=1,2,3$. The analysis is carried out by viewing the string Ansatz of Eq. \eqref{Standard string ansatz} as the background field configuration, which we denote as $\bar{\Psi}$. We then take into account small perturbations on top of the background configuration as $\Psi = \bar{\Psi} + \epsilon \delta \Psi$, where $\delta \Psi = \{ \delta h, \delta h^\dagger, \delta \phi^\pm, \delta\phi^3, \delta A_\mu^\pm, \delta A_\mu^3 \}$ and $\epsilon$ is a (infinitesimally) small book-keeping factor. The general idea of the analysis is to substitute the perturbed fields $\Psi$ in the EoMs listed above, and expand up to first order in $\epsilon$. This will lead to coupled linear differential equations in the perturbations: $\mathcal{D}\delta \Psi = 0$. The stability can then be examined by considering small time oscillations of frequency $\omega$ as in  $\delta \Psi = \delta \psi(\rho, \theta) e^{i\omega t}$. 
Upon the choice of an ansatz for $\delta\psi$, the linear matrix equations will result in Schrödinger-like equations $\mathcal{D}\delta \psi = \omega^2 \delta \psi $ for the frequency squared $\omega^2$. If an imaginary eigenvalue $\omega^2 < 0$ is found in the spectral decomposition of $\mathcal{D}$, the string ansatz is classically unstable. On the other hand, if no negative $\omega^2$ is found, the string is concluded to be classically stable.

\paragraph{The stability sectors} 
Carrying out the procedure just discussed, we can now substitute the perturbed background field configuration $\Psi = \bar{\Psi} + \epsilon \delta \Psi$ in the equations of motion \eqref{EOM Full own model in h}-\eqref{EOM Full own model in A}. Collecting only the terms that are $\mathcal{O}(\epsilon)$ results in a linear system of equations in the perturbations, $\mathcal{D}\delta \Psi = 0$. The symmetries of the background string allow for a decomposition of the original equations in four distinct blocks:
\begin{align}
    \mathcal{D}_{1}\,\begin{pmatrix} \delta h_{1}  \\ \delta h_1^{\dagger} \\ \delta A^{3}_\mu \end{pmatrix} = 0, && \mathcal{D}_{2}\,\delta \phi^3 = 0, 
     \\
     \mathcal{D}_{3} \begin{pmatrix} \delta h_2  \\ \delta \phi^{-}  \\ \delta A_\mu^{-} \end{pmatrix} = 0,
     \label{eq:D3}
    && \mathcal{D}_{3}^{\dagger} \begin{pmatrix} \delta h_2^{\dagger}  \\ \delta \phi^{+}  \\ \delta A_\mu^{+} \end{pmatrix} = 0.
\end{align}
In what follows, we shall investigate the stability of the string in each of these sectors through a systematic overview of the elements of their differential operators. 

\paragraph{Gauge fixing} 
In order to simplify the expressions to come, and to remove all non-physical degrees of freedom ahead in the field perturbations, we propose, based on \cite{Goodband:1995rt, Goodband:1995he}, the following $R_\xi$-gauge fixing choice on the gauge field perturbations:
\begin{align}
    F_{1}\left(A^{-}_{\mu}\right) &= \partial_{\mu} \delta A^{-\mu} + ig \bar{A}^{3}_{\mu}\delta A^{-\mu} -\frac{ig}{\sqrt{2}}\bar{h}_1^\dagger\delta h_2 - ig \bar{\phi}^3  \delta \phi^-= 0,\label{Gauge choice AMIN}\\
    F_{2}\left(A^{+}_{\mu}\right) &= \partial_{\mu} \delta A^{+\mu} - ig \bar{A}^{3}_{\mu}\delta A^{+\mu} +\frac{ig}{\sqrt{2}}\bar{h}_1\delta h_2^\dagger + ig \bar{\phi}^3  \delta \phi^+= 0, \label{Gauge choice APLUS}\\
    F_{3}\left(A^{3}_\mu\right) &= \partial_{\mu}\delta A^{3\mu} - \frac{ig}{2}\left(\bar{h}_{1}^{\dagger} \delta h_{1} - \bar{h}_{1} 
 \delta h_{1}^{\dagger}\right) = 0. \label{Gauge choice metastable A3}
\end{align}

\subsection{The ghost degrees of freedom} 
Not all linear perturbations $\delta \Psi$ correspond to physical fluctuations. The gauge redundancy of the theory makes that certain linear perturbations correspond to infinitesimal gauge transformations of the field. As such, these directions need to be studied separately, and a consistent gauge theory should contain no unstable ghost fluctuation mode. The fields of the theory transform under an infinitesimal gauge transformation according to 
\begin{align}
    \delta h &= \frac{i}{2}\delta\theta^a(x) \sigma^a \bar{h}\\
    \delta \phi^a &= - \epsilon^{abc}\delta\theta^b(x) \bar{\phi}^c \\
    \delta A_\mu^a &= \frac{1}{g}\partial_\mu \delta\theta^a(x) - \epsilon^{abc}\delta\theta^b(x) \bar{A}_\mu^c,
\end{align}
where the functions $\delta \theta^a(x)$ characterize the gauge transformation. The gauge fixing of Eqs. \eqref{Gauge choice AMIN}-\eqref{Gauge choice metastable A3} does not fix the gauge completely. There are still residual degrees of freedom  which need to be investigated.
This is done by substituting the linear field perturbations as those from the infinitesimal transformations in the adequate sectors. Since the fields $\delta A_\mu^-$ and $\delta A_\mu^+$ are each others complex conjugate, it is sufficient to check the stability from the gauge fixing of $A^3$ and $A^-$ only.

Starting with the $A^3$ fluctuations, we find after considering linear time fluctuations $\delta\theta^3\to\delta\theta^3 e^{i\omega t}$ that the ghost perturbation satisfy the following Schrödinger-like equation:
\begin{equation}
    \left(-\nabla^2 + \frac{g^2}{2}|\bar{h}_1|^2\right)\delta \theta^3 = \omega^2 \delta \theta^3.
\end{equation}
Numerical computations of the eigenvalue spectrum shows that all $\omega^2$ are positive. Therefore this equation yields no instability, confirming the consistency of the analysis.

Doing the same for the ghosts resulting from the gauge fixing of $\delta A^-_\mu$ one finds the following stability equation for $\delta\theta^- \equiv (\delta\theta^1 + i \theta^2)/\sqrt{2}$:
\begin{equation}
    \left(-\nabla^2 + 2ig \bar{A}^{3k}\partial_k + g^2 \bar{A}^{3k}\bar{A}^{3k} + g^2 |\bar{h}_1|^2 + g^2 V^2\right)\delta\theta^- = \omega^2 \delta\theta^-
\end{equation}
Similarly, this ghost stability equation contains no unstable modes. We conclude that none of the residual gauge freedoms cause an instability.

\subsection{Instabilities of the local string sector ($h_1, h_1^\dagger, A_\mu^3$)}
We start by analyzing the sector of the fields ($h_1, h_1^\dagger, A_\mu^3$) with stability operator $\mathcal{D}_1$, which only depends on $\beta_S$. 
It turns out that this sector is the same as the one for a standard local $U(1)$ string in the abelian Higgs model. We refer to \cite{Goodband:1995rt} for a more exhaustive analysis of this sector. Since the abelian string is topological, the unit-winding string $n=1$ is absolutely stable regardless of the values of $\beta_S$.

\subsection{Stability from the $\phi^3$ sector}
The sector for $\mathcal{D}_2$ contains a single linear equation for $\delta\phi^3$ which simply reads
\begin{equation}
    \left( \Box + \frac{\gamma}{2}\left|\bar{h}_1\right|^2 + 8\tilde{\lambda}V^2 \right)\delta\phi^3 = 0.
    \label{LSA Eq Full model dphi3}
\end{equation}
To make further progress, we assume a cylindrical  ansatz on the perturbation $\delta\phi^3$ with small fluctuations in time:
\begin{equation}
    \delta\phi^3(t,\rho, \theta) = a(\rho) e^{il\theta}e^{i\omega t}.
\end{equation}
Substituting this form in Eq. \eqref{LSA Eq Full model dphi3}, and rescaling according to \eqref{Rescale of coordinate in Low E theory} with $\omega^2 \to \tilde{\omega}^2 = \omega^2 / m_Z^2$ and $a\to Va$ yields the following eigenvalue equation
\begin{equation}
    \left[ -\frac{d^2}{d\rho^2} - \frac{1}{\rho} \frac{d}{d\rho} + \frac{l^2}{\rho^2} + \frac{\alpha}{\eta} f^2(\rho) + \beta_S \eta  \right]a(\rho) = \omega^2a(\rho).
\end{equation}
Since the centrifugal term stabilizes the string, it is sufficient to analyze the case $l=0$ of minimal centrifugal force. In addition, in virtue of having a positive definite potential for every values of the parameters, we conclude that a fluctuation in the $\delta\phi^3$ field about the string configuration does not probe an unstable eigenvector direction with $\omega^2 < 0$. This agrees with our numerical findings and we thus conclude that the $\mathcal{D}_2$ sector contains no unstable direction.

\subsection{Instabilities in the ($h_2, \phi^-, A^-_\mu$) sector}
Since the sector defined by $\mathcal{D}_3$ and $\mathcal{D}_3^\dagger$ are each others conjugates, it is sufficient to consider only one of them. We choose the former.  After fixing the gauge freedom by imposing Eqs. \eqref{Gauge choice AMIN} and \eqref{Gauge choice APLUS}, due to the $t$- and $z$-independence of the background fields, the stability equations of this sector in \eqref{eq:D3} further decouple into:
\begin{equation}
    \begin{pmatrix}
        D_1 & X & Y_i \\ 
        X^\dagger & D_2 & 0\\
        Y^\dagger_j & 0 & D_{{3}_{ji}}
    \end{pmatrix}
    \begin{pmatrix}
        \delta h_2 \\ \delta \phi^- \\ \delta A^{-\,i}
    \end{pmatrix} = 0
    \label{D2 Eqn 1}
\end{equation}
and 
\begin{equation}
    \left(-\nabla^2 + 2ig\bar{A}^{3,k}\partial_k + g^2 \bar{A}^{3,k} \bar{A}^{3,k} + \frac{1}{2}g^2 |\bar{h}_1|^2 + g^2V^2 \right)\begin{pmatrix}
        \delta A^-_t \\\delta A^-_z 
    \end{pmatrix} = 0.
\end{equation}
The latter equations yield no instability and so we focus solely on the first set of equations \eqref{D2 Eqn 1}. The corresponding matrix elements are listed below. We recall that in our metric conventions of mostly minus, the positions of the spatial indices are relevant. We find:
\begin{align*}
    D_1 &= -\nabla^2 +ig\bar{A}^{3,k}\partial_k + \frac{1}{4}g^2 \bar{A}^{3,k}\bar{A}^{3,k} + \gamma V^2 + 2\lambda \left(|\bar{h}_1|^2 - v^2\right) + \frac{1}{2}g^2 |\bar{h}_1|^2, \\
    D_2 &= - \nabla^2 + 2ig\bar{A}^{3,k}\partial_k + g^2 \bar{A}^{3,k} \bar{A}^{3,k} + \frac{1}{2}\gamma |\bar{h}_1|^2 + g^2V^2, \\
    D_{{3}_{ji}} &= -\eta_{ji}\left(-\nabla^2 + 2ig \bar{A}^{3,k}\partial_k +g^2 \bar{A}^{3,k}\bar{A}^{3,k} + \frac{1}{2}g^2|\bar{h}_1|^2 + g^2V^2 \right) - 2ig \bar{A}^{3}_{ji}, \\
    X &= \frac{V}{\sqrt{2}}\left( g^2 - \gamma \right)\bar{h}_1, \\ 
    Y_{i} &= -\sqrt{2}ig \left( \, \partial_{i} - \frac{ig}{2} \bar{A}^3_i \right) \bar{h}_{1},
\end{align*}
where $\eta_{ji}$ is the spatial metric which in our convention is $\eta_{ji}=-\delta_{ji}$, and we defined $\bar{A}^{3}_{ij} \equiv \partial_i \bar{A}^3_j - \partial_j \bar{A}^3_i$. In these equations we already substituted the background field configuration of $\bar{\phi}^3 = V$ as it simplifies the matrix by making the off-diagonal elements relating $\phi^-$ and $A^-_\mu$ to vanish.

\paragraph{Expanding the gauge fields in the spin basis} To make an ansatz on the perturbations, it is convenient to change basis and write the component of the gauge field $\delta A_\mu^-$ perturbations in their spin basis:
\begin{equation}
    \delta A^\pm_{\uparrow} = \frac{e^{-i\theta}}{\sqrt{2}}\left(\delta A^{\pm}_\rho - i \delta A^{\pm}_\theta \right), \qquad \delta A^\pm_{\downarrow}  = \left(\delta A^\mp_{\uparrow} \right)^\dagger = \frac{e^{i\theta}}{\sqrt{2}}\left(\delta A^{\pm}_\rho + i \delta A^{\pm}_\theta \right).
\end{equation}
In this basis, the perturbations correspond to eigenstates of the total angular momentum operator $\hat{J}_z = \hat{L}_z + \hat{S}_z$ since these operators act as:
\begin{equation}
    \hat{L}_z = - i\frac{d}{d\theta}, \qquad (\hat{S}_zA^\pm)_j = -i \epsilon_{3jk}A_k^\pm,
\end{equation}
respectively. It then follows that the gauge fields in the spin basis satisfy: $(\hat{S}_zA^\pm)_\uparrow = + A^\pm_\uparrow$ and $(\hat{S}_zA^\pm)_\downarrow = - A^\pm_\downarrow$  such that $\delta A^-_\uparrow$ is the spin-up component and $\delta A^-_\downarrow$ is the spin-down component of the gauge field \cite{Goodband:1995he}.

\paragraph{Perturbation ansatz and final equations}
To obtain a usable form for the stability equations in the perturbations, we make the following Fourier mode expansion on all of the perturbations by decomposing the perturbations in a basis of eigenstates of the orbital angular momentum operator $\hat{L}_z$:
\begin{align}
    \delta h_2(x) &= e^{i\omega t}\sum_m s_m(\rho) e^{im\theta}\supset s_l(\rho)e^{il\theta}e^{i\omega t} \label{Ansatz 1 perturbation full model},\\
    \delta \phi^-(x) &= e^{i\omega t}\sum_{m} a_m(\rho)e^{im\theta} \supset a_{l-n}(\rho)e^{i(l-n)\theta} e^{i\omega t}, \label{Ansatz 2 perturbation full model} \\
    \delta A^{-}_{\uparrow} &= e^{i\omega t}\sum_m iw_{\uparrow m}(\rho)e^{im\theta} e^{-i\theta} \supset iw_{\uparrow\, l-n}(\rho)e^{i(l-n-1)\theta} e^{i\omega t},\\
    \delta A^{-}_{\downarrow} &= e^{i\omega t}\sum_m -iw_{\downarrow m}(\rho)e^{im\theta}e^{i\theta} \supset -iw_{\downarrow\, l-n}(\rho)e^{i(l-n+1)\theta} e^{i\omega t}, \label{Ansatz 4 perturbation full model}
\end{align}
where we selected a specific combination of Fourier modes which make the resulting linear equations of motion real. Finally, we substitute these ansatz into the stability equations obtained in Eq.  \eqref{D2 Eqn 1}, and we rescale the coordinate and fields according to:
\begin{equation}
    \rho \to \tilde{\rho} = m_Z\rho, \quad (s, a,w_\uparrow, w_\downarrow) \to v(\tilde s, \tilde{a}, \tilde w_\uparrow, \tilde w_\downarrow)
\end{equation}
to, in the end, obtain the following set of matrix eigenvalue equation:
\begin{equation}
    \begin{pmatrix}
        D_{1} & A & B & C \\
        A & D_{2} & 0 & 0 \\
        B & 0 & D_{3} & 0\\
        C & 0 & 0 & D_{4}
    \end{pmatrix} \begin{pmatrix}
        s_l \\ a_{l-n} \\ w_{\uparrow\, l-n }\\ w_{\downarrow\, l-n}
    \end{pmatrix} = \omega^2 \begin{pmatrix}
        s_l \\ a_{l-n} \\ w_{\uparrow\, l-n }\\ w_{\downarrow\, l-n}
    \end{pmatrix}
    \label{Eq Eigenvalue equation full model}
\end{equation}
where the elements of this stability matrix are given by:
\begin{align}
    D_{1} &= -\frac{d^2}{d\rho^2} - \frac{1}{\rho} \frac{d}{d\rho} + \frac{\left(l+n\zeta(\rho)\right)^2}{\rho^2} + \frac{\beta_S}{2}\left(f^2(\rho)-1\right) + f^2(\rho) + \alpha \label{D1 stability eqn} \\
    D_{2} &= -\frac{d^2}{d\rho^2} - \frac{1}{\rho}\frac{d}{d\rho} + \frac{(l-n+2n\zeta(\rho))^2}{\rho^2} + \frac{\alpha}{\eta} f^2(\rho) + \eta \label{D2 stability eqn} \\
    D_3 &= -\frac{d^2}{d\rho^2} - \frac{1}{\rho}\frac{d}{d\rho} + \frac{\left(l-1-n+2n\zeta(\rho) \right)^2}{\rho^2} + \frac{4n}{\rho} \frac{d\zeta(\rho)}{d\rho} + f^2(\rho) + \eta \label{D3 stability eqn} \\
    D_4 &= -\frac{d^2}{d\rho^2} - \frac{1}{\rho}\frac{d}{d\rho} + \frac{\left(l+1-n+2n\zeta(\rho) \right)^2}{\rho^2} - \frac{4n}{\rho} \frac{d\zeta(\rho)}{d\rho} + f^2(\rho) + \eta \label{D4 stability eqn} \\
    A &= \sqrt{\eta}\left(1-\frac{\alpha}{\eta}\right) f(\rho) \label{A stability eqn} \\
    B &= \sqrt{2}\left(\frac{df(\rho)}{d\rho} - \frac{n}{\rho}f(\rho)(1-\zeta(\rho))\right) \label{B stability eqn} \\
    C &= -\sqrt{2} \left( \frac{df(\rho)}{d\rho} + \frac{n}{\rho} f(\rho) (1-\zeta(\rho)) \right)  \label{C stability eqn}
\end{align}
For simplicity, we dropped the tildes, but the equations above should be understood as being dimensionless.
We note that the stability equations depend on the fundamental parameters listed in Eq. \eqref{Fundamental parameters definition}, with the exception of $\beta_M$.
We solve these equations numerically and investigate the classical stability of the string solutions by  determining the eigenvalue spectrum of this matrix operator, and extract the lowest eigenvalue $\omega^2$. 
The equations are solved by using the string profiles of Section \ref{Section String Profile}, and by turning the coupled differential equation into a finite-dimensional matrix equation with the appropriate boundary conditions.

\paragraph{Boundary conditions}
All field configurations satisfy a Dirichlet boundary condition at spatial infinity: $\delta \psi(\rho)=0$ as $\rho\to\infty$.
On the other hand, in the origin, the boundary condition depends on the considered perturbation mode labeled by the integers $l\in\mathbb{Z}$. Different boundary conditions may apply on the field perturbations for different decay modes. From ansatz \eqref{Ansatz 1 perturbation full model}-\eqref{Ansatz 4 perturbation full model} on the perturbations, we can choose the boundary conditions to ensure regularity at the origin. These suggest the following boundary conditions:
\begin{align}
    s_l(\rho): \quad    &\begin{cases}
	 \text{Dirichlet: }\: s_l(0)=0 & \text{when }\: l \neq 0 \\
	 \text{Neumann: }\: s_l'(0)=0 & \text{when }\: l=0
                \end{cases} \\
    a_{l-n}(\rho): \quad    &\begin{cases}
	 \text{Dirichlet: }\: a_{l-n}(0)=0 & \text{when }\: l \neq n \\
	 \text{Neumann: }\: a_{l-n}'(0)=0 & \text{when }\: l=n
                \end{cases} \\
    w_{\uparrow l-n}(\rho): \quad    &\begin{cases}
	 \text{Dirichlet: }\: w_{\uparrow l-n}(0)=0 & \text{when }\: l \neq n+1 \\
	 \text{Neumann: }\: w_{\uparrow l-n}'(0)=0 & \text{when }\: l=n+1
                \end{cases} \\
    w_{\downarrow l-n}(\rho): \quad    &\begin{cases}
	 \text{Dirichlet: }\: w_{\downarrow l-n}(0)=0 & \text{when }\: l \neq n-1 \\
	 \text{Neumann: }\: w_{\downarrow l-n}'(0)=0 & \text{when }\: l=n-1
                \end{cases}
\end{align}
The various centrifugal terms above suggest to consider in particular the modes $l=0, \pm1, \cdots,\pm(n-1), \pm n, \pm(n+1)$ for winding number $n$. This is because as we increase the mode $|l|$, the centrifugal terms become dominant and make the string more stable. Hence, it is sufficient to look at the modes $|l|$ which make the centrifugal contributions minimal, as they will lead to an enlarged instability region. In particular, this means that for a unit winding string, the considered modes should be $l=0,\pm1, \pm2$. 

\paragraph{Eigenfunction normalization} 
In what follows and in Appendix \ref{Appendix LSA} we will inspect the properties of the eigenfunction associated to the unstable direction that can arise from the eigenvalue equation in Eq. \eqref{Eq Eigenvalue equation full model}.
Given a perturbation mode $l$, the four component eigenvector 
$\delta \Psi_l = \{s_l, a_{l-n}, w_{\uparrow l-n}, w_{\downarrow l-n} \}$ is normalized with the 
two-dimensional norm as
\begin{align}
\label{eq:normalization}
    \braket{\delta \Psi_l, \delta \Psi_l} &\equiv \int d\theta \int\rho\, d\rho\, \delta\Psi^\dagger_l \delta\Psi_l \\
    &= 2\pi \int\rho\, d\rho \left(s_l(\rho)^2 + a_{l-n}(\rho)^2 + w_{\uparrow l-n}(\rho)^2 + w_{\downarrow l-n}(\rho)^2 \right) \equiv 1
    \nonumber
\end{align}
The relative importance of each component in the eigenfunction can then be evaluated by computing the relative norm using the same measure. In particular, it can be used to tell in which direction instabilities are aligned the most.

\section{Classical stability and impact for PTAs}\label{Section numerical results}

By solving numerically the eigenvalue problem presented in the previous section, we now turn to determine the
conditions for the classical stability
of the string solution reviewed in Section \ref{Section String Profile}. We also discuss the impact of our results for the metastable string interpretation of the PTA signal.

The classical instability can arise in the sector involving the coupled linear fluctuations of the fields $\{ h_2, \phi^{-},A^{-} \}$, namely in the eigenvalue equation \eqref{D2 Eqn 1} and \eqref{Eq Eigenvalue equation full model}.
This stability sector only depends on the parameters 
$\alpha,\beta_S$ and $\eta$; three of the four parameters listed in Eq. \eqref{Fundamental parameters definition}.
For the unit winding string $n=1$, we observe that the $l=\pm1, \pm2$ perturbation modes yield no instabilities. In the following analysis, we thus only consider the $l=0$ mode for which we do observe instabilities.

\begin{figure}
    \centering    \includegraphics[width=10cm]{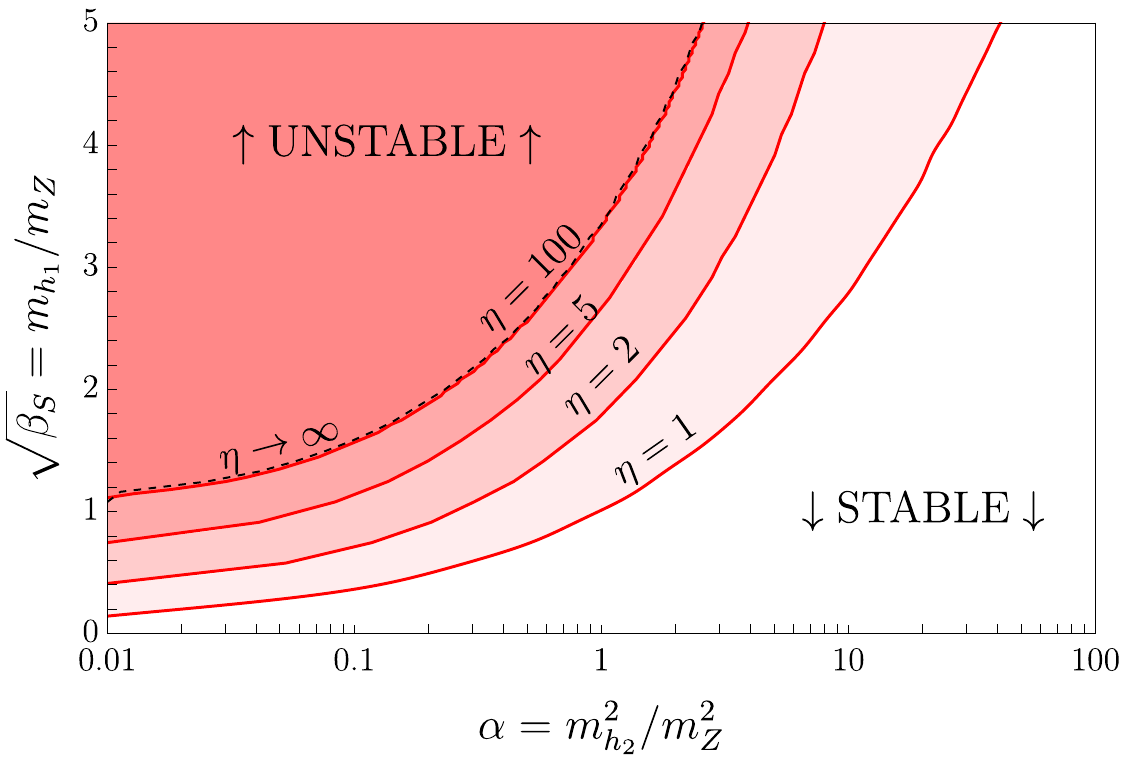} 
    \caption{Classical stability analysis of the model parameter space with $\sqrt{\beta_S} = m_{h_1} /m_Z$ and $\alpha = (m_{h_2}/m_Z)^2$ for different values of the hierarchy of scales $\eta = (m_W/m_Z)^2$. The red regions indicate those in which the string ansatz of Eq. \eqref{Standard string ansatz} is classically unstable. The dashed line corresponds to the stability line in the limit of $\eta\to\infty$ where the adjoints and $A^\pm$ bosons are frozen (see section \ref{section eta to inf} for a derivation). 
    }
    \label{Fig LSA Full Model own params}
\end{figure}

The results of our analysis are presented in Figure \ref{Fig LSA Full Model own params}, where we identify the regions in the ($\alpha$, $\sqrt{\beta_S}$) parameter space for which the strings are classically unstable, for different values of $\eta$. 
In general, 
instabilities arise for low values of the portal parameter 
$\alpha$ and large values of $\beta_S$.
A small value of the portal coupling corresponds to a small effective mass term for the second doublet component, see \eqref{eq:effectivePot}, favoring an instability along such direction in the field fluctuations.
An instability is induced also by a large $\beta_S$, which  controls the size of a negative term in the differential operator $D_1$, see \eqref{D1 stability eqn}.

The instability region gets larger when reducing the hierarchy between the two symmetry breaking scales controlled by $\eta$, as it becomes easier for the field fluctuations to explore the $SU(2)$ structure of the vacuum manifold. 
As we increase $\eta$, the stability lines converge to a single curve.
In fact, in the limit $\eta\to\infty$ the differential operator of the eigenvalue problem in \eqref{Eq Eigenvalue equation full model} simplifies, and the 
eigenvalue problem can be reduced to a single stability equation for the fluctuation of $h_2$, 
as we show in Appendix \ref{section eta to inf}.
The instability line of this simplified problem is shown as a dashed line in Figure \ref{Fig LSA Full Model own params} for comparison.
We also note that in the limit of $\alpha\to0$ and $\eta\to\infty$, the string is classically stable for $\beta_S\leq1$ and unstable for $\beta_S>1$. This corresponds to the semi-local string limit; see e.g. \cite{Vachaspati:1991dz, Achucarro:1999it} for a review as well as Appendix \,\ref{App electroweak string}.

In all the parameter space that we have explored, 
there is only one negative eigenvalue in the unstable region.
We now study the property of the corresponding eigenvector (see Appendix \ref{Appendix LSA} for more details on the eigenvalue spectrum and the eigenfunction shapes).
The direction associated to the negative eigenvalue 
involves the four-mode expansion $\delta\Psi_l = \{s_l(\rho), a_{l-n}(\rho), w_{\uparrow l-n}(\rho), w_{\downarrow l-n}(\rho) \}$ in \eqref{Eq Eigenvalue equation full model} corresponding to the fluctuations of the fields 
$\{\delta h_2,\: \delta \phi^-,\: \delta A_\mu^-\}$.
It is interesting to investigate which of the four components dominates in the unstable direction by
computing the relative norm fraction of the different components via the norm definition in Eq. \eqref{eq:normalization}. 
This ratio can indicate if the instability is driven primarily by the scalar doublet, or instead if it is driven by the triplet or gauge fields. We consider the same parameter space as above and  compute the relative norm fraction of all the four components.
By inspection, we find that the dominant contributions to the eigenfunction are always given by the mode $s_l(\rho)$ of $\delta h_2$ or $w_{\downarrow l-n}(\rho)$ of $\delta A_\downarrow$, while the other two components are subleading in the total norm contribution.

\begin{figure}[t!]
    \centering
    \includegraphics[width=7.3cm]{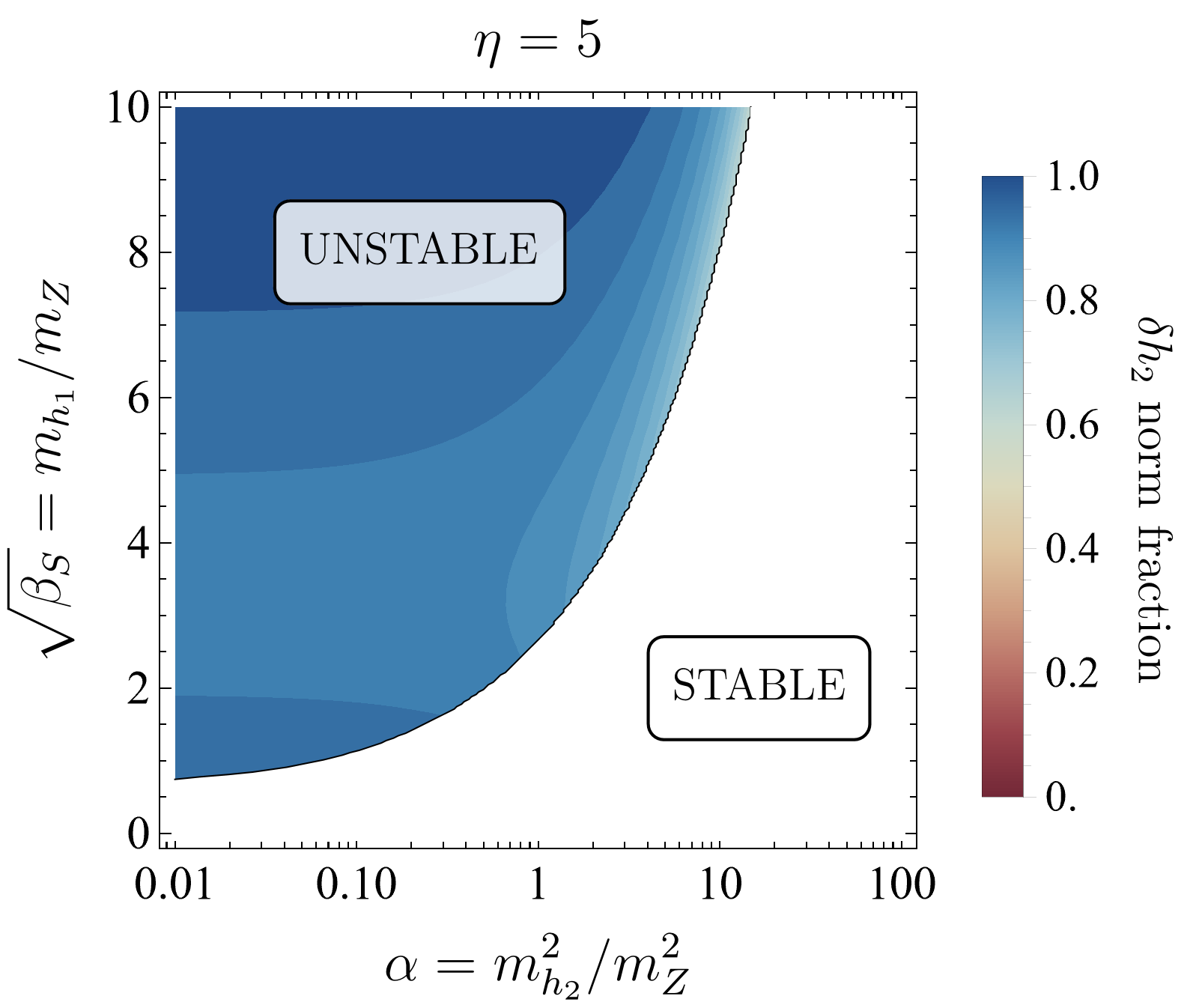}
    \includegraphics[width=7.3cm]{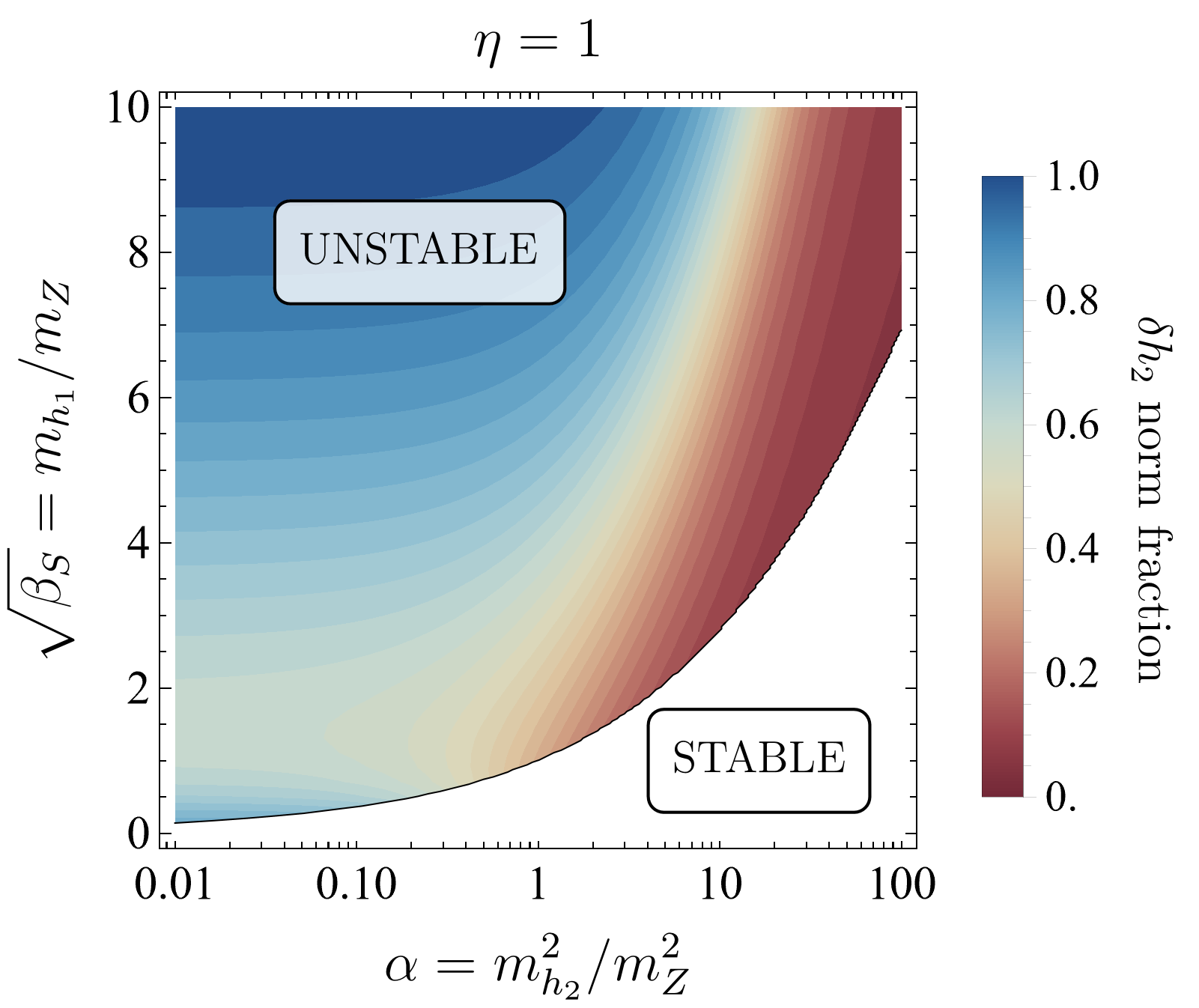}
    \caption{The norm fraction $\braket{s_{l=0}, s_{l=0}}/\braket{\delta \Psi_{l=0}, \delta \Psi_{l=0}}$ in the unstable regions for $\eta=5$ (on the left), and $\eta=1$ (on the right) where $s_l$ is the $\delta h_2$ fluctuation. Blue regions indicate domination by the $\delta h_2$ perturbations, while red regions are dominated by $\delta A_\downarrow$ via $w_{\downarrow l-n}$. 
    For a larger hierarchy $\eta$, the instability is almost entirely dominated by $\delta h_2$, while a lower hierarchy allows for domination by $\delta A_\downarrow$.}
    \label{Fig L2 Domination Plot}
\end{figure}

In Figure \ref{Fig L2 Domination Plot}
we then
plot, in the unstable region, the size of the relative $\delta h_2$ norm.
In the blue region the 
eigenfunction is dominated by 
the mode 
$s_l(\rho)$ of $\delta h_2$, while in the red region
by the mode
$w_{\downarrow l-n}(\rho)$ of $\delta A_\downarrow$.
For large $\eta$ the instability is mainly driven by the $\delta h_2$ fluctuation. For small $\eta$, instead, not only does the instability region get larger, but the extra unstable region is dominated by the gauge field fluctuations leading to a gauge field condensation in the string core\,\footnote{A similar analysis can be carried out for the electroweak string for varying values of $\sin \theta_W$. It indicates that for $\sin \theta_W \to 1$ the instability is dominated by the Higgs field, while for $\sin \theta_W \to 0$ it is dominated by the $W$ bosons leading to 
$W$-condensation. We revisit this in Appendix \ref{App electroweak string}.} as visible from the eigenfunctions in Appendix \ref{Appendix eigenvalues&functions}.

\medskip

\paragraph{Impact on PTAs} Given the classical stability analysis presented above, we can now investigate its implication for the metastable string interpretation of the PTA signal.
The classical stability of the string solution depends on the parameters $\alpha, \beta_S$ and $\eta$.
When metastable, the string life time generically depends on the ratio of the monopole mass over the string tension, which in our parametrization \eqref{Fundamental parameters definition} can be written as
\begin{equation}
    \kappa = \frac{m^2}{\mu} = \frac{4\pi}{g^2}\frac{f_M^2(\beta_M)}{f_S(\beta_S)} \eta,
    \label{Eq kappa in function of params}
\end{equation}
where $f_M$ is the numerical monopole mass function defined in Eq.\,\eqref{Numerical function monopole mass} and $f_S$ is the numerical string tension function introduced earlier in Eq.\,\eqref{Numerical function string tension}. 

\begin{figure}
    \centering
\includegraphics[width=0.49\textwidth]{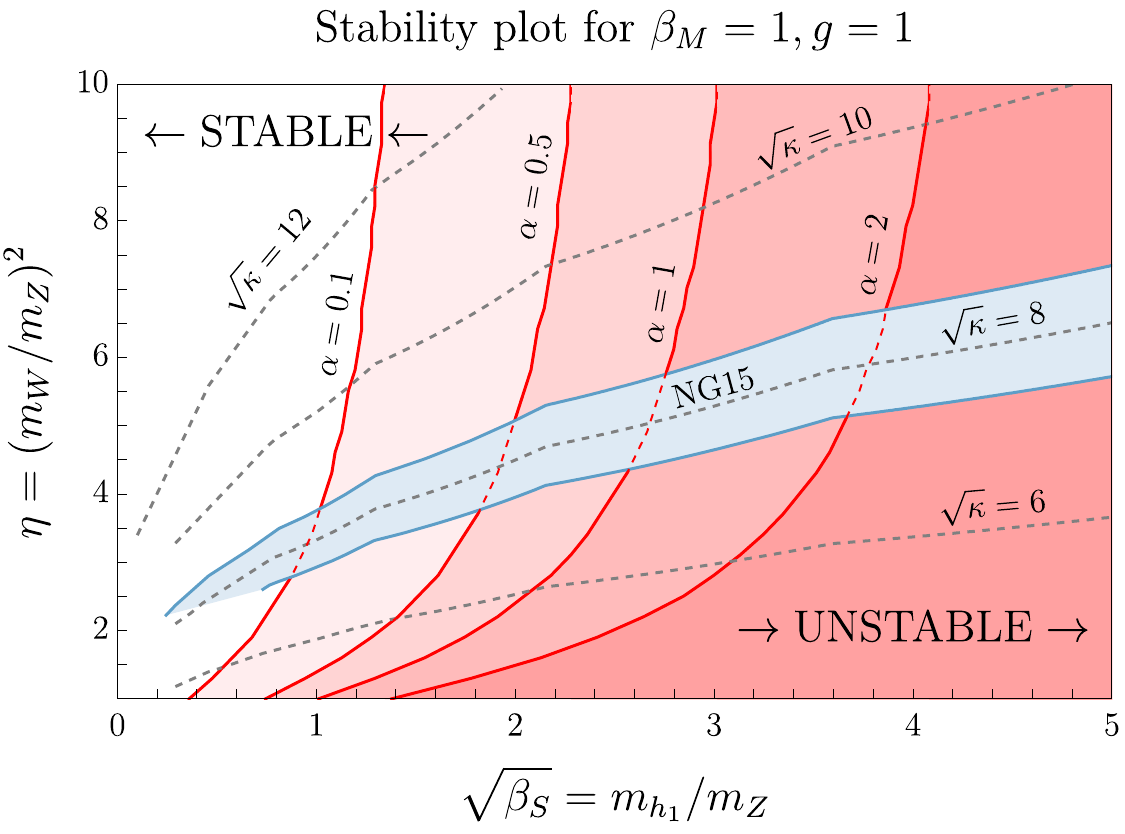}
    \includegraphics[width=0.49\textwidth]{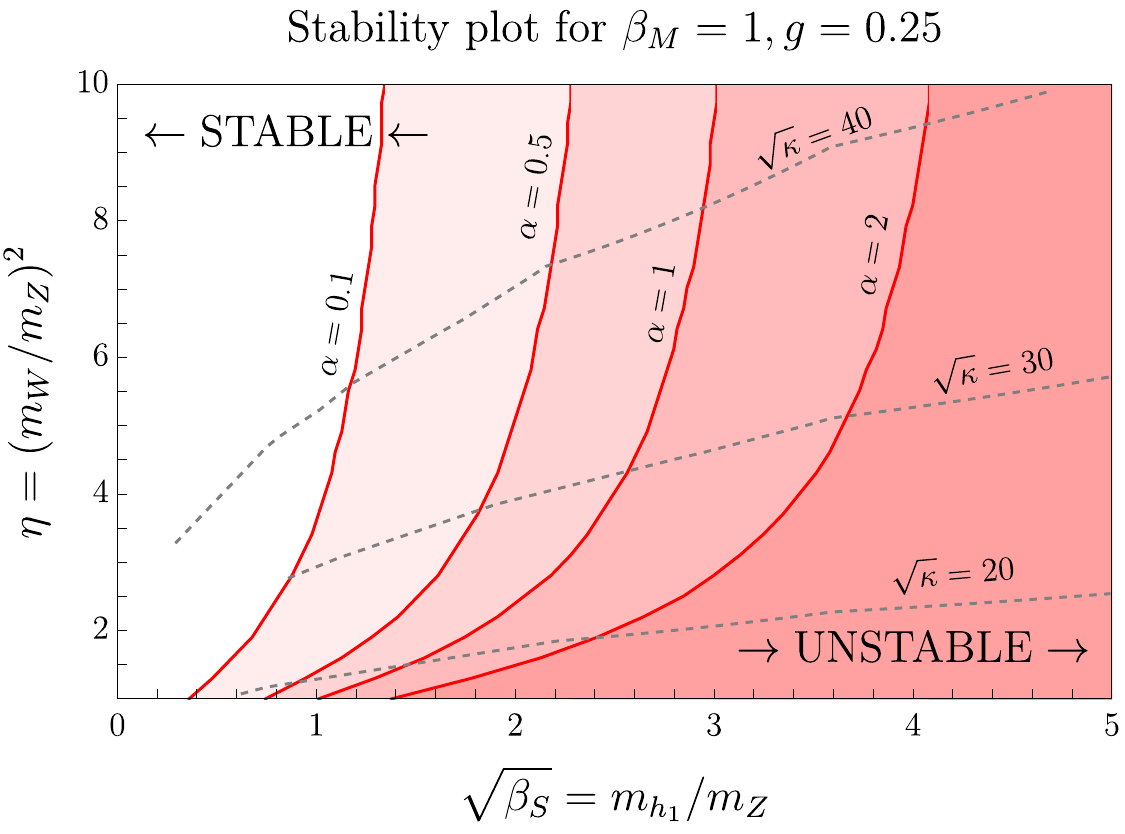}
    \caption{The bounds from the stability analysis in the $(\sqrt{\beta_S}, \eta)$-plane for different values of the portal parameter $\alpha$. The (overlapping) red regions indicate the regions in the parameter space for which the string is unstable. \textbf{The top panel} shows it in the case where $\beta_M =1$ and $g=1$ with corresponding isocurves of $\sqrt{\kappa}$. The preferred PTA region ($\sqrt\kappa \in [7.5,\, 8.5]$) is also indicated. \textbf{The bottom panel} has been done in the case where $\beta_M =1$ and $g=0.25$, thus accommodating for larger values of $\sqrt{\kappa}$ via Eq. \eqref{Eq kappa in function of params}. }
     \label{Fig:money_plots}
\end{figure}

The classical instability is independent of $\beta_M$ and the parameter $\kappa$ mildly depends on $\beta_M$, so we fix $\beta_M=1$ in what follows.
However, the monopole mass depends strongly on the gauge coupling, so we consider two benchmarks for illustration, one with weak gauge coupling ($g=0.25$) and one with strong gauge coupling ($g=1$).
The parameter $\kappa$ is then mainly controlled by the value of $\eta$, and does not depend on $\alpha$.
We henceforth display our results in the $\eta$ vs $\beta_S$ plane with varying values of $\alpha$, in Figure \ref{Fig:money_plots}.
The region on the right of a given contour with fixed $\alpha$ is classically unstable.
Classical stability is generically favored for larger $\alpha$, while it is disfavored for larger $\beta_S$ and smaller $\eta$.

In the left panel of Figure \ref{Fig:money_plots}, we show the case of $g=1$ which leads to relatively small values of $\sqrt{\kappa}$ in this parameterization. In this plot, we highlight in light blue the region $\sqrt{\kappa} \simeq 8$ as favored by PTAs when assuming infinite long strings and the thin wall approximation for the quantum tunneling (see  Section \ref{Section Metastabel Strings and GWs}). As we can see, depending on the value of $\alpha$, string solutions that would otherwise explain the PTA signal are actually classically unstable.

In the right panel of Figure \ref{Fig:money_plots}, we show the case of weak coupling, $g=0.25$. Here, the values of $\sqrt{\kappa}$ are generically larger than in the left panel for the region of cosmological interest with $\eta > 1$. Such values of $\sqrt{\kappa}$ would not be compatible with the PTA signal according to the analysis of \,\cite{NANOGrav:2023hvm}, but may become viable when considering the additional aspects pointed out in \,\cite{Tranchedone:2026lav}. Even in this case with $\sqrt{\kappa} \gg 1$, we observe the onset of classical instability in a large part of the parameter space depending on $\alpha$.

Overall, we find that next to the regions of classical stability, a significant portion of the model parameter space relevant for PTAs can actually lead to classically unstable strings. In the next section, we discuss the possible phenomenological implications of such instability.

\section{Following the instability}\label{Section chasing the instability}

The stability analysis presented in the previous section shows the presence of unstable fluctuations around the string background in regions of parameter space that are (or can be) relevant for the metastable string interpretation of the PTA signal. 
In those regions, the commonly adopted analysis of section \,\ref{Section Metastabel Strings and GWs} needs to be revised. However, the phenomenological implications of such instability can be very different depending on how it actually develops.
The exponential growth of fluctuations around the string can in fact have two very different outcomes:
\begin{enumerate}
\item The initial string profile completely unwinds upon classical evolution, with the final state being a homogeneous field configuration in the vacuum manifold.
The cosmic string network dissolves on a time scale much shorter than the Hubble time soon after formation, leading to negligible gravitational wave emission.
\item The growth of the fluctuations stops upon reaching a new classically stable string profile, and the network survives. Such new configuration will generally differ from the original ANO string, in particular due to the presence of scalar condensates in the core and a decreased string tension. Still, this new string will be metastable and presumably decay via monopole nucleation.
The analysis of section \,\ref{Section Metastabel Strings and GWs} may still apply, albeit with a different tension and tunneling rate.
\end{enumerate}

To distinguish between these two possibilities one necessarily needs to go beyond the fluctuation analysis. A possible strategy would be to study the fate of the ANO string in the low-energy theory where only the Higgs doublets and $A^3_\mu$ are dynamical.
At leading order, the Lagrangian is actually the one given in \eqref{eq:partforstring}. Within such a theory, any instability of the ANO string will eventually be stabilized into a new string configuration with a condensate for $h_2$ over distances $1/m_{h_2}$ around the string core, realizing the second outcome above.
However, this result holds trivially as we are neglecting from the start any correction coming from the heavy $SU(2)$ states. Therefore, while this may be considered as a hint in favor of a new string profile resolving the instability, there is no conclusive evidence that the string will not rather unwind due to the dynamics of the additional heavy degrees of freedom. 

The way to investigate the fate following the instability would then be to solve the full system in Eq.\,\eqref{EOM Full own model in A} in time and determine whether the initial string configuration eventually unwinds or stabilizes itself, which is however numerically demanding. Therefore, to shed light on this matter, we follow the instability in a toy model where the gauge interactions are switched off and which is much easier to investigate. This toy model of course has massless degrees of freedom (Goldstone bosons) that would not appear in the full theory. Such states would become heavy when restoring gauge interactions, and we expect the classically-unstable global string solutions to be more likely to unwind in the global case.

\begin{figure}
\centering
\includegraphics[width=10cm]{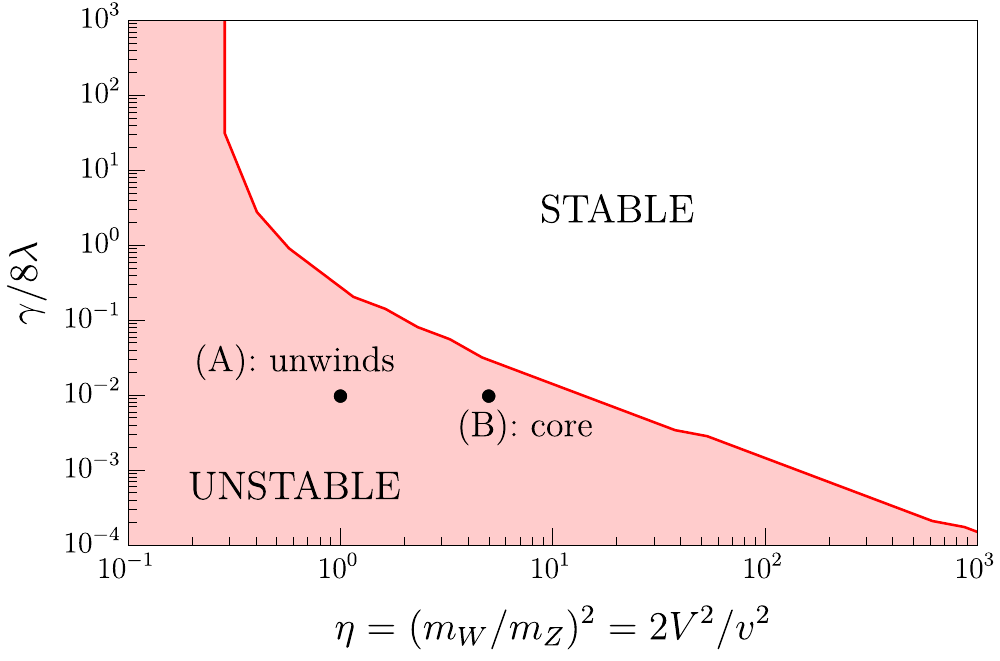}
 \caption{Stability analysis of the global model obtained by setting the gauge fields to zero in Eq. \eqref{Lagrangian full theory OWN model}. Notice that the full parameter space can now be covered by two independent variables. The two considered benchmark points (A) and (B) are also indicated.
   \label{fig:lsa_global_model}
 }
  \end{figure}

We then consider the theory in Eq.\,\eqref{Lagrangian full theory OWN model} and set the $SU(2)$ gauge fields and coupling to zero.
The classical stability analysis for this toy model is shown in Fig.\,\ref{fig:lsa_global_model}.
We then start with a perturbed global string configuration for benchmarks where it is classically unstable, and evolve it in time according to the full equations of motion with both a staggered leapfrog method and a relaxation algorithm on a two-dimensional grid; both methods giving the same results. See Appendix \ref{Appendix Global Model} for details on the stability analysis and on the time evolution.
We fix portal and quartic couplings to representative values:
\begin{equation}
    \frac{\gamma}{8\lambda} = 0.01, \quad \frac{\tilde\lambda}{\lambda} = 1,
\end{equation}
and study two different scenarios depending on the ratio of the symmetry breaking scales:
\begin{equation}
    \text{(A)}: \,\, \eta=2 V^2/v^2 = 1, \quad 
    \text{(B)}: \,\, \eta=2V^2/v^2 = 5.
\end{equation}
We show these benchmark points in the stability plot of Fig.\,\ref{fig:lsa_global_model} for convenience.

\begin{figure}
    \centering
\includegraphics[width=\linewidth]{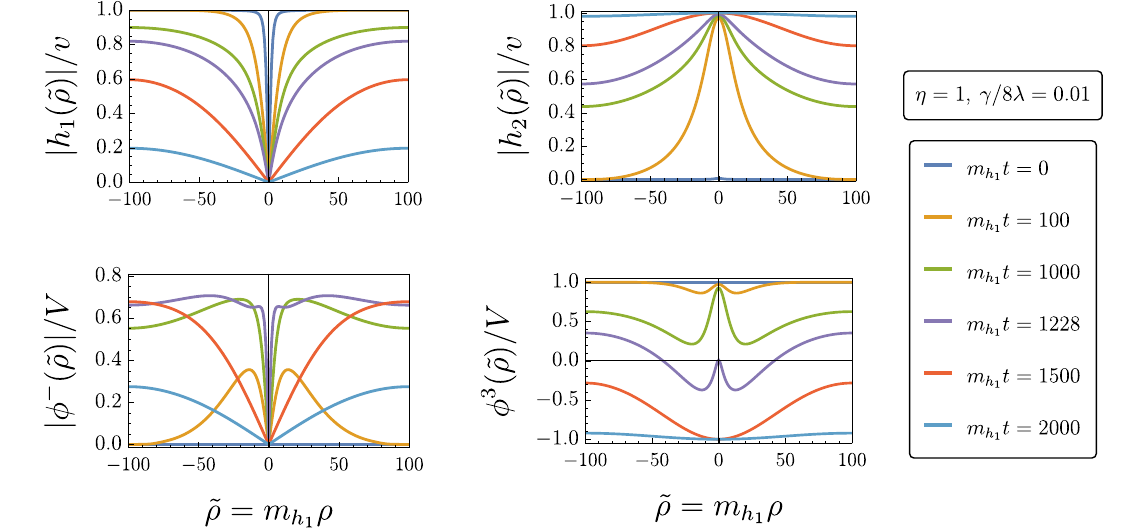} 
    \caption{Time evolution of the profiles as a function of the cylindrical radius $\tilde\rho$ for the unwinding string benchmark with $\eta = 2V^2/v^2 = 1$ and $\gamma/8\lambda=0.01$. For this benchmark we considered a 2D cartesian grid defined on the interval $[-100, 100]^2$ with as uniform grid spacing given by $\Delta \tilde x= \Delta \tilde y = 1.0$ and time steps $\Delta \tilde t=0.1$.   }
    \label{Fig Unwinding BM Time Evolution}
\end{figure}

\begin{figure}
\centering
    \includegraphics[width=10cm]{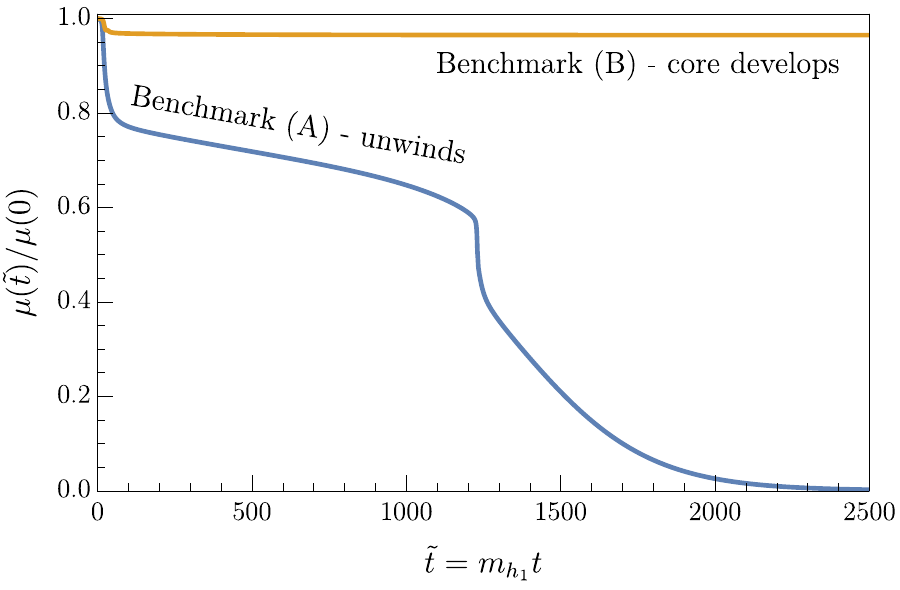}
 \caption{Evolution of the tension $\mu$ as a function of time. A clear distinction is observed between the two chosen benchmarks. Benchmark (A), for which $V<v$, has a clear monotonically decreasing tension which ultimately vanishes. On the other hand, for benchmark (B) with $V>v$, the tension initially decreases but soon comes to a halt. We conclude that (A) leads to a full dissipating string, while (B) leads to a non-decaying string with a non-trivial condensate in its core.
   \label{fig:tension_evolution}
 }
  \end{figure}

\begin{figure}
\centering
    \includegraphics[width=\linewidth]{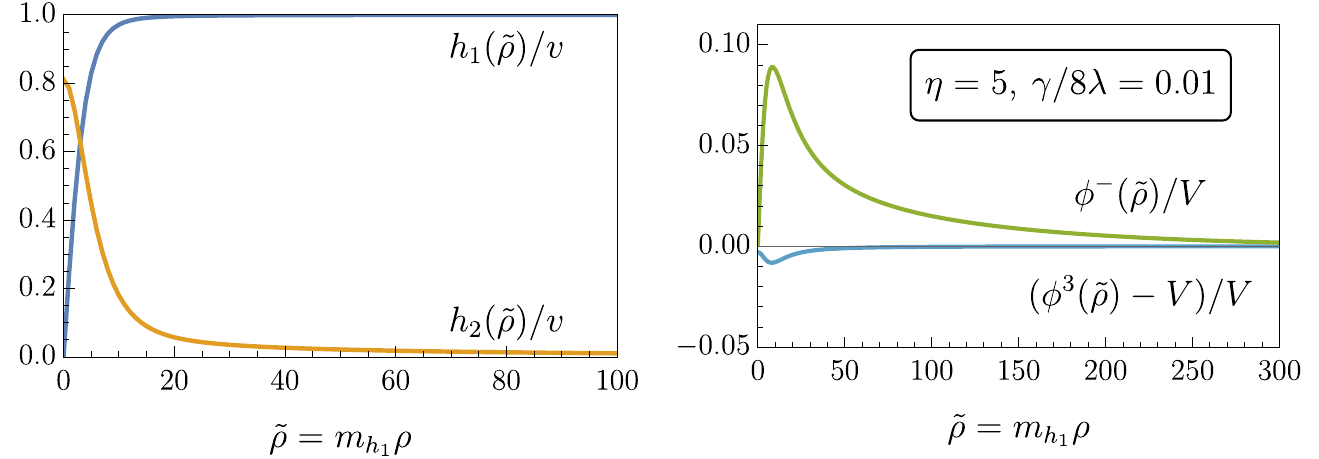}
 \caption{Profiles of the benchmark point (B) with $\eta=2V^2/v^2 = 5$ and $\gamma/8\lambda = 0.01$ as a function of the cylindrical radius $\tilde\rho$. We observe that all fields $h_2$, $\phi^-$ and $\phi^3$ develop a non-zero condensate in the core of the string. For this benchmark we considered a 2D cartesian grid defined on the interval $[-500, 500]^2$ with as uniform grid spacing given by $\Delta \tilde x= \Delta \tilde y = 1.0$ and time steps $\Delta \tilde t=0.1$.  
 }
   \label{fig:core_string}  
  \end{figure}

Our numerical results are shown in Fig.\,\ref{Fig Unwinding BM Time Evolution} for the benchmark point (A) and in Fig.\,\ref{fig:core_string} for (B). 
As we can see, (A) leads to a complete unwinding of the initial (global) string. This appears to occur via the growth of $h_2$, which then sources $\phi^\pm$ via the portal interaction with a string-like profile enforcing $\phi^\pm=0$ at the core. Once these profiles are large enough, namely $\mathcal{O}(V)$, they induce some pull on the radial component of the triplet $\phi^3$. Eventually, the value of $\phi^3$ at the center of the string crosses zero and flips sign, $\phi^3(\tilde \rho=0) = -V$. At this point, the string can unwind by simply expanding the $h_2$ core, thus ending up with $h_2 = v$ and $h_1=0$ everywhere in space. In the core region, $\phi^\pm$ can now relax to zero as well, leaving no defect behind. The string tension as a function of the relaxation time for this benchmark point is shown in Fig.\,\ref{fig:tension_evolution} (blue line). As we can see, the string decay proceeds without encountering any barrier, namely the string tension monotonically decreases, and is therefore purely classical\,\footnote{Notice that here the string tension eventually goes to zero due to the finite size of the box. In the realistic scenario, this would correspond to the location of a neighboring string.}. 
By looking at the tension, we can also notice the two--step nature of this process. At the very beginning, the tension decreases due to the formation of the doublet core. The winding is then transferred to the triplet components, until the the dynamics of $\phi^3$ becomes important around $\tilde t \sim 1000$ (see also Fig.\,\ref{Fig Unwinding BM Time Evolution}) and initiates the complete decay.

When repeating the same analysis for the benchmark point (B), we instead find that the growth of $\phi^\pm$ around the string actually stops, and the $h_2$ core attains a stationary size. We show the late-time string configuration in Fig.\,\ref{fig:core_string}.
We observed that the $\phi^\pm$ profiles tend to keep on evolving at infinity, which we suspect to be because of the global nature of the string. In practice, we stop our algorithm when the relative change on the fields is below $10^{-6}$.
These results suggest that a string with a locally stable core may indeed be possible in the global theory.

The qualitatively different behavior of the benchmark points (A) and (B) may be understood as follows. The initial string configuration starts with tension $\mu \sim v^2$. In order to grow the triplet fields to the point where they can backreact on $\phi^3$, one needs profiles of $\mathcal{O}(V)$. Such gradients correspond to energy per unit length $\sim V^2$. Therefore, this first step is precluded when $V > v$ independently of the other model parameters.

We can summarize the results of this section as follows. While we only provided a simplified study of the theory of interest without gauge interactions, we find evidence for the formation of a string core following the instability instead of full unwinding for $V \gtrsim v$,
which is the parameter space of interest for the cosmological evolution considered in section \,\ref{Section Metastabel Strings and GWs}. We may expect this behavior to persist in the full theory given that gauging will lift the Goldstone directions of our toy model, even though a detailed investigation will be needed to clarify this open question.
A string profile with a core would still imply a possibly large emission of GWs from the metastable strings, but would also require a novel study of the tunneling process in the classically unstable regions of Fig.\,\ref{Fig:money_plots} to assess the life time of the network and its implications for PTAs.

\section{Conclusion}
\label{sec:conclusion}

Metastable cosmic strings are a promising candidate for explaining the gravitational-wave background observed at PTAs. Contrary to stable strings, the network has a finite life time which implies a IR cut-off for the gravitational wave spectrum. This in turn allows for a better fit to the PTA data. In the simplest scenario, the IR cut-off is controlled by the rate of monopole-antimonopole nucleation, a process that leads to the decay of the entire string network thus halting the gravitational wave emission thereafter.

In this paper, we have performed a comprehensive analysis of the classical stability of the string solutions arising in the simplest $SU(2) \rightarrow U(1) \rightarrow 1$ model for metastable strings\,\cite{Shifman:2002yi} customarily considered in the literature in terms of the underlying microscopical parameters. In fact, classical stability is a necessary condition for identifying the correct initial state for the quantum tunneling into monopoles. Classically unstable strings would then invalidate the standard picture reviewed in section \,\ref{Section Metastabel Strings and GWs}.

Our main results are presented in Figure \,\ref{Fig:money_plots}. As we can see, there exist regions of the parameter space favored by PTAs where the strings are indeed classically stable, and the analysis of section \ref{Section Metastabel Strings and GWs} applies. However, next to these regions we identify portions of parameter space where strings that would otherwise be a viable explanation of the PTA signal are actually classically unstable, invalidating the standard treatment of section \ref{Section Metastabel Strings and GWs}.
This can occur for moderate but also for large monopole mass over string tension ratio $\kappa$, as we show in the right panel of Fig. \ref{Fig:money_plots}.

Classically unstable strings may either completely dissolve soon after formation, therefore leading to negligible gravitational wave emission, or instead reach a new configuration where the field profiles rearrange themselves to decrease the string tension. Determining which of these two options is actually realized in the parameter space of interest requires to go beyond the fluctuation analysis carried out in this paper. Based on a simplified version of the metastable string model of\,\cite{Shifman:2002yi} where the gauge interactions are switched off, we nevertheless find some evidence for a new string solution to be realized following the classical instability. We leave the study of the full model including gauge interactions for future work.

If a new string solution is indeed approached following the classical instability, the PTA signal may be explained even in the classically unstable part of the parameter space in Figure \,\ref{Fig:money_plots}. This would however require a new analysis for determining the rate of quantum tunneling as well as the new string tension, which could induce a sizable change in the time of the network decay.

\paragraph{Acknowledgment}
We thank Yu Hamada for discussions, and A\"{a}ron Rase for collaboration in the initial stage of this project. SB is supported by the Deutsche Forschungsgemeinschaft under Germany’s Excellence Strategy---EXC 2121 ``Quantum Universe"---390833306.
AM is supported in part by the Strategic Research Program and Large Research Group High-Energy Physics of the Research Council of the Vrije Universiteit Brussel, and
by the FWO project FWO-IRI I000725N ``Empowering Tomorrow's Technological Horizons for the Einstein Telescope".
MG is supported by FWO-Vlaanderen through grant number 1166626N.

\begin{appendices}

\section{Review of GW from metastable strings}
\label{app:reviewGWmetastable}
This appendix serves as an outline of the main steps involved in the computation of the GWB from (metastable) cosmic strings, which were presented in Figure \ref{Fig GW spectrum from metastable strings}. Our discussion is primarily based on the more detailed reviews in \cite{NANOGrav:2023hvm, Auclair:2019wcv, Gouttenoire:2019kij, Blasi:2020mfx, Servant:2023tua,Schmitz:2024gds}.

Gravitational waves from cosmic strings are in dominant part generated by loops of cosmic strings through the presence of small scale structures. The physical properties of loops in the network are therefore the central quantities that build up the GWB generated by the network. Each loop is characterized by its size $l_k$ at birth, which is determined by the frequency of standing waves present on the loop. Taking into account the redshift from the expansion of the Universe, the size of the loops is labeled by a positive integer $k$:
\begin{equation}
    l_k = \frac{2k}{f}\frac{a(t)}{a(t_0)}.
\end{equation}
Roughly speaking, the gravitational wave background relies on two extra ingredients: the stochastic distribution of loops $n(l_k, t)$ as a function of time in the network and the average power emitted by a single loop. Using these quantities, the gravitational wave background from a network of stable cosmic strings can be expanded in different power harmonics labeled by $k$:
\begin{equation}
     \Omega_{\text{GW}}(f) = \frac{f}{\rho_c} \frac{d\rho_{\text{GW}}}{df} = \sum_{k=1}^\infty \Omega_{\text{GW}}^{(k)}(f), 
     \label{Eq GWB CS Harmonic Expansion}
\end{equation}
where $f$ is the frequency and $\rho_c = 3 H_0^2 / (8 \pi G_N)$ is the critical energy density of the Universe. The contributions to the GWB from the different wave numbers are found to be 
\begin{equation}
     \Omega^{(k)}_{\text{GW}}(f) = G\mu^2 \frac{f}{\rho_c} C_k(f) \Gamma_k ,
\end{equation}
where $\Gamma_k = \Gamma k^{-q}/\zeta(q)$ is a dimensionless factor related to the partially radiated power through gravitational waves. Here $\Gamma$ is a dimensionless coefficient appearing in the total radiated power in the form of gravitational waves: $P\approx \Gamma G\mu^2$. It is numerically found to be $\Gamma\approx 50$. Lastly, the spectral index $q$ varies with the considered microscale structures on the string loops. Since cusps are the dominant channel of GW emission, we take throughout our computations $q=4/3$ only. The different coefficients $C_k(f)$ encode the density of loops in the network, as well as the evolution of the network via:
\begin{equation}
     C_k(f) = \frac{2k}{f^2}\int_{t_F}^{t_0} dt\, \Theta(t) \left( \frac{a(t)}{a(t_0)} \right)^5 n(l_k,t) ,
\end{equation}
where $t_F$ is the time of formation of the network, which we take to be $10^{-30}$ s, and where the loop number density is given by 
\begin{equation}
     n(l_k, t) = \frac{\mathcal{F}}{t_k^4}\left( \frac{a(t_k)}{a(t)}  \right)^3 \frac{C_{\text{eff}}}{\alpha(\alpha+\Gamma G\mu)}, 
\end{equation}
for $\mathcal{F}\approx 0.1$ a numerical correction factor and $\alpha \approx 0.1$ another numerical parameter for the average size of loops at formation. Furthermore, the time parameter $t_k(t)$ is the time at which loops formed at time $t$ start to contribute to the GW emission and is given by  
\begin{equation}
    t_k = \frac{l_k/t + \Gamma G\mu}{\alpha + \Gamma G\mu}t.
\end{equation}
$C_{\text{eff}}$ effectively captures the epoch of the Universe: $C_{\text{eff}}\approx 5.4$ for radiation domination and $C_{\text{eff}}\approx 0.39$ for matter domination. The Heaviside function $\Theta(t)$ is inserted in order to ensure a proper evolution of the loop dynamics. Namely, it enforces that the loops are formed before today ($t_k < t_0$), that the time of loop formation happens after the birth of the network ($t_k>t_F$) and finally that loops are formed at a size $l_k = \alpha t$. Therefore,
\begin{equation}
    \Theta(t) = \theta(t_0-t_k)\theta(t_k - t_F)\theta(\alpha - l_k/t).
\end{equation}
Lastly, we also note that in practice the infinite series in Eq. \eqref{Eq GWB CS Harmonic Expansion} needs to be truncated to some upper $k$ value. We take $k_{\text{max}} = 10^6$. Besides, the summation for large values of $k$ can be approximated by the continuum limit of its sum:
\begin{equation}
    \sum_{k=m}^{k=n}\Omega_{\text{GW}}^{(k)}(f) \approx f^{1-q} \int_{f/n}^{f/m} dx\, x^{q-2} \Omega^{(1)}_{\text{GW}}(x)
\end{equation}
In our computations, we compute the sum for the first values of $k$ up to 100. Afterwards, we use the latter integral approximation up to $k_{\text{max}}$. This concludes how to model the GWB from a network of stable cosmic strings. 

In the case of metastable cosmic strings, one needs to modify the previous expressions in order to capture the effect of (late time) string annihilation via the nucleation of monopoles \cite{NANOGrav:2023hvm, Buchmuller:2023aus}. In practice, this can be achieved in a rather minimal way by adding an exponential suppression to the loop number density and add another Heaviside function as:
\begin{equation}
    n^{\text{metastable}}(l_k, t) = \theta(\tau_d - t_k) E(l_k, t) n(l_k, t).
\end{equation}
The Heaviside function phenomenologically ensures that the string loop production is no longer efficient once the loop formation time $t_k$ exceeds the decay time $\tau_d$ of strings via monopole annihilation. This decay rate $\tau_d$ can be computed from Eq. \eqref{eq:thin_wall} by $\tau_d = \Gamma_d^{-1/2}$. The factor $E(l_k, t)$ effectively models the exponential suppression of loops contributing to the GWB:
\begin{equation}
    E(l_k, t) = \exp\left( -\Gamma_d \left[ l_k(t-t_k) +\tfrac{1}{2}\Gamma G\mu (t-t_k)^2 \right] \right)
\end{equation}
To get the spectrum of the metastable strings, it is then sufficient to replace the loop number density $n(l_k, t)$ with the newly modified one $n^{\text{metastable}}(l_k, t)$ in the previous expressions.

\section{More details on the stability analysis}\label{Appendix LSA}
With the main results of the stability analysis presented in Section \ref{Section numerical results}, it is useful to consider several different aspects we did not consider in the main part of the text. In particular, we hereby present the evolution of the lowest eigenvalues and different benchmarks of the corresponding eigenfunctions in Section \ref{Appendix eigenvalues&functions}. In addition, we discuss in more details the instability arising in the limit of an infinite hierarchy $\eta\to \infty$ in Section \ref{section eta to inf} dominated by $\delta h_2$ and of the instability dominated by the $A^\pm$ gauge fields in \ref{Appendix low eta limit}. Finally, we also present a comparison of our results with another source in the literature \cite{Chitose:2023dam} in \ref{Appendix comparison Chitose}. 

\subsection{Evolution of eigenvalues and eigenfunctions}\label{Appendix eigenvalues&functions}
\begin{figure}[t]
    \centering
    \includegraphics[width=7.3cm]{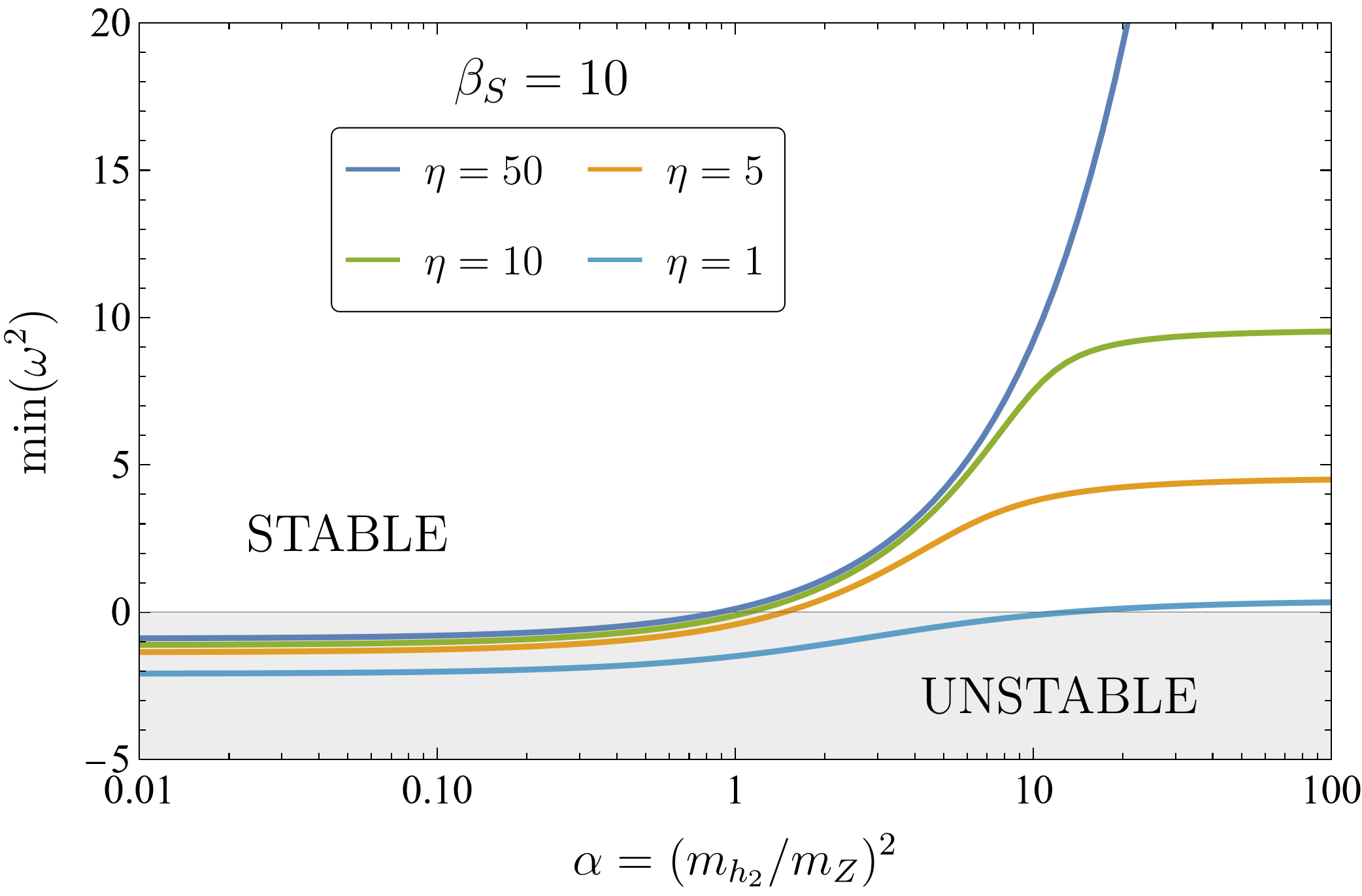}
    \includegraphics[width=7.3cm]{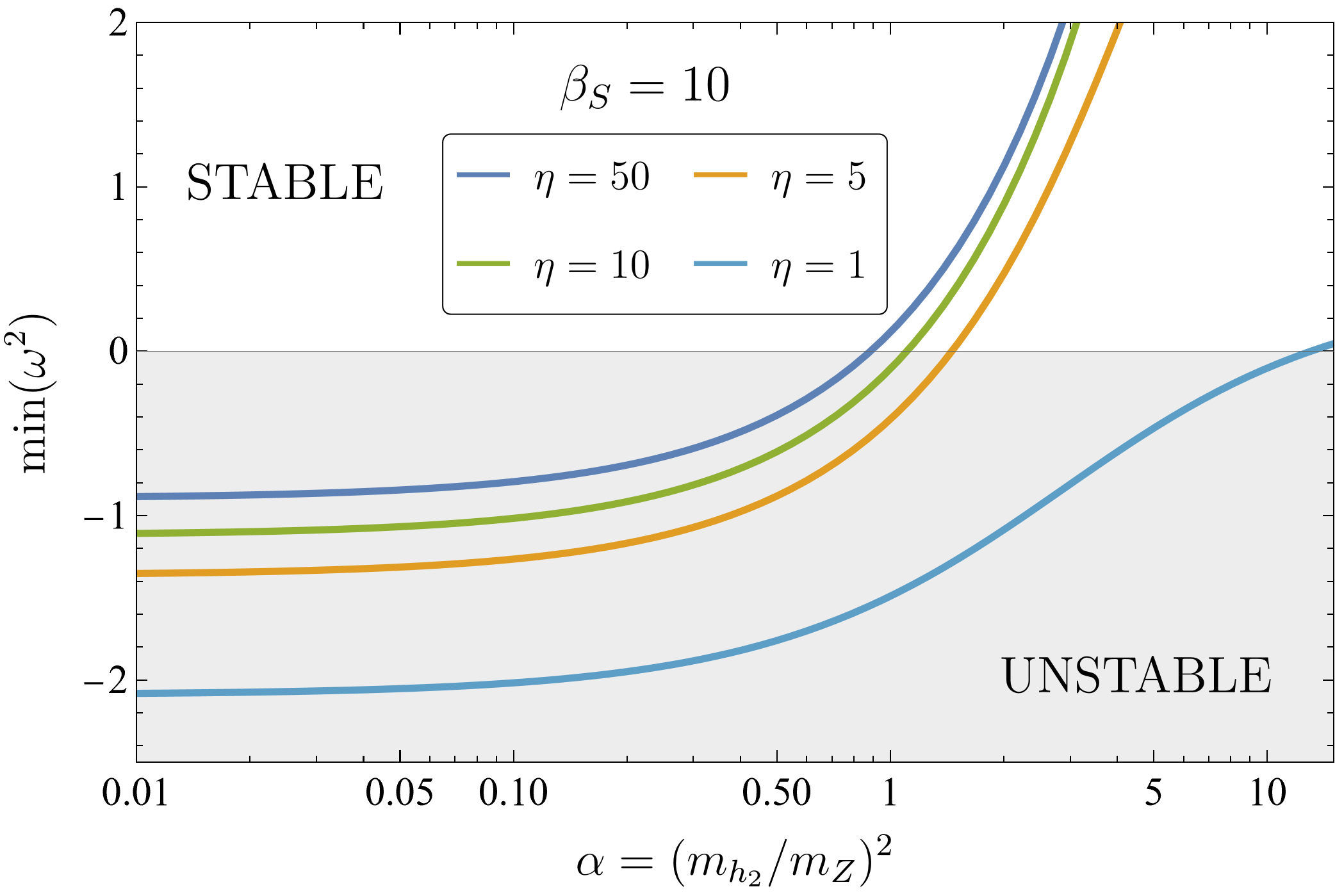}
    \caption{Evolution of the lowest eigenvalue for a string of winding $n=1$ and $\beta_S=(m_{h_1}/m_Z)^2=10$ fixed, as a function of $\alpha=(m_{h_2}/m_Z)^2$ and of the hierarchy of scales $\eta = (m_W/m_Z)^2$. The right panel is a zoomed-in portion of the left panel.}
    \label{Fig:eigenvalues}
\end{figure}

To study the evolution of the lowest eigenvalues we take a slice in the parameter space by taking a fixed value of $\beta_S$ and by looking at how the eigenvalues vary as a function of the portal parameter $\alpha$. In order to lie deeper in the unstable region, we fix $\beta_S=10$. The evolution of the lowest eigenvalues are reported in Figure \ref{Fig:eigenvalues} for different values of $\eta$.

We again observe two different ways an instability can arise, as consistent with what we previously discussed via Figure \ref{Fig L2 Domination Plot}. 
On one hand, we observe that the eigenvalues are almost independent on $\alpha$ in the instability region of small $\alpha < \beta_S$ and larger $\eta$ values. In this regime the contribution from the Mexican hat potential for the fundamental scalars dominates over the portal contribution (which induces a mass to $h_2$ in the low-energetic theory). Accordingly, the instability is dominated by the $\delta h_2$ perturbation which seeds the formation of a condensate in the central region of the string. 
On the other hand, another type of instability is primarily induced by lowering $\eta$. Therefore, this instability corresponds to excitations in the gauge fields which similarly grows a $A^\pm$-condensate in the core analogous to the $W$-condensate instability of the electroweak string (see Appendix \ref{App electroweak string}). 

\begin{figure}[h!]
    \centering
    \includegraphics[width=\linewidth]{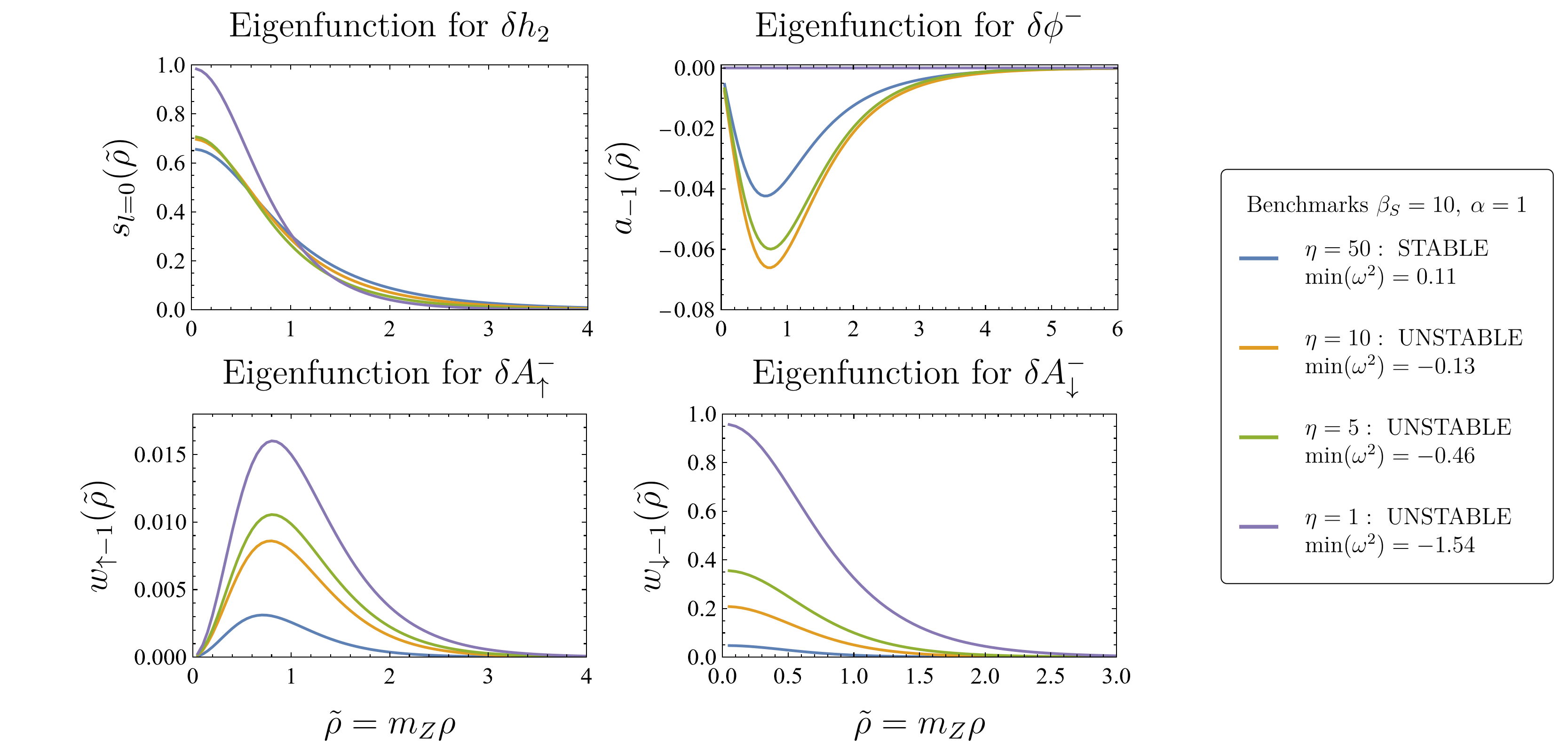}
    \caption{Benchmarks of eigenfunctions corresponding to the lowest eigenvalue for fixed values of $\beta_S = 10$ and $\alpha=1$ for decreasing values of the hierarchy $\eta$. We observe an increasing domination of the $A^-$ perturbations in the core as we lower the hierarchy.}
    \label{Fig Eigenfunctions}
\end{figure}

To confirm the latter distinction, we investigate the shape of the corresponding eigenfunctions for various benchmarks, which are normalized via the norm defined in Eq. \ref{eq:normalization}. We do this for the same benchmark $\beta_S = 10$ but with $\alpha=1$ fixed as well. This is presented in Figure \ref{Fig Eigenfunctions}. In this Figure it is once again manifest how the perturbation transitions from a $\delta h_2$ dominated profile to one dominated by the  gauge boson (specifically $\delta A_{\downarrow}^-$) as we reduce the hierarchy $\eta$. 
We observe how all the perturbations are located around/near the origin indicating that all types of instabilities arise from the core and grow outwards (whether the instability continues growing or stops to form a finite-sized condensate is another question we addressed in Section \ref{Section chasing the instability} in the case of the global model). 
Finally, we note that in Figure \ref{Fig Eigenfunctions} we display the eigenfunctions also for stable benchmarks where the smallest eigenvalue is actually positive\footnote{The eigenvalue problem we solve is analogous to a quantum mechanical problem where the asymptotic value of the potential energy is shifted, and hence it may admit bound state solutions localized around the origin also when the eigenvalue is zero or positive.}.

\subsection{Limit of infinite hierarchy $\eta\to\infty$}\label{section eta to inf}
The $h_2$ dominated instability can be traced back to the same instability that arises in the case of semi-local strings \cite{Vachaspati:1991dz, Achucarro:1999it}. This instability can essentially be isolated by taking the infinite hierarchy limit: $\eta\to\infty$ for $\alpha$ and $\beta_S$ finite. In this limit, the stability matrix in the direction of the gauge bosons (Eq. \eqref{D3 stability eqn}-\eqref{D4 stability eqn}) essentially traces out, and any instability, if any, will come from the reduced eigenvalue equation: 
\begin{equation}
    \begin{pmatrix}
        D_1 & A\\
        A & D_2
    \end{pmatrix}
    \begin{pmatrix}
        s_l \\
        a_{l-n}
    \end{pmatrix}
    = \omega^2 \begin{pmatrix}
        s_l \\
        a_{l-n}
    \end{pmatrix}
\end{equation}
with $D_1$, $D_2$ and $A$ still given by Eqs. \eqref{D1 stability eqn}, \eqref{D2 stability eqn} and \eqref{A stability eqn}, respectively. We can use this simplified system to analyze the limit of $\eta\gg1$ more carefully. In this limit, one can show that $s_l$ and $a_{l-n}$ are related via
\begin{equation}
    a_{l-n}(\rho) = \frac{-f(\rho)}{\sqrt{\eta} \left( 1-\omega^2 /\eta \right) } s_l(\rho).
\end{equation}
Substituting this back into the eigenvalue equation for $s_l$, the instability problem is reduced to the single eigenvalue equation:
\begin{equation}
    \left[ -\frac{d^2}{d\rho^2} - \frac{1}{\rho}\frac{d}{d\rho} + \frac{(l+n\zeta(\rho))^2}{\rho^2} + \frac{\beta_S}{2}(f^2(\rho)-1) + \alpha \right]s_l(\rho) = \omega^2 s_l(\rho).
    \label{Stability equation eta to infinity}
\end{equation}
This is, apart from the extra portal coupling $\alpha$, the same instability equation found in a similar stability analysis for semi-local strings \cite{Vachaspati:1991dz, Achucarro:1999it}. This observation already leads us to certain expectations of outcome. For instance, in the limit where $\alpha\to 0$ (portal term vanishes), the instability equation is the exact same as that of the semi-local string. In this section of parameter space, an extra global $SU(2)$ symmetry is enhanced and therefore the string solutions are non-topological and no longer guaranteed to be classically stable. It is then known that the, now non-topological, $n=1$ string solution is dynamically stable only when $\beta_S \leq 1$, but unstable once $\beta_S >1$. The corresponding stability equation of Eq. \eqref{Stability equation eta to infinity} is solved numerically, and the result is presented in Figure \ref{Fig LSA Full Model own params} as the dotted black curve. As a consistency check, we note that in the limit $\alpha\to 0$ the semi-local limit is consistently satisfied, as mentioned previously. 
Physically, this instability is driven by the minimization of the potential energy. Neglecting the gradient energy, this means that it occurs in the rough limit of $\beta_S \gtrsim 2\alpha$. Of course this is only a rough stability bound since the gradient energy (of the scalar and gauge field) contributes positively, and makes the string more stable but, at least, it explains qualitatively the semi-local limit instability.

\subsection{Instability from $A^\pm$ gauge condensation}\label{Appendix low eta limit}
The gauge dominated instability of the string can be traced back by looking at the effective mass acquired by the $A^\pm$ gauge fields. We first note that the energy density is given by
\begin{multline}
    \mathcal{H} = -(D_i h)^\dagger D^i h - \frac{1}{2}(D_i\phi^a)(D^i\phi^a)  + V(h^\dagger h) + \frac{1}{2} G_{ij}^+ G^{-ij} + \frac{1}{4}Z_{ij}Z^{ij}  \\ 
    - ig Z^{ij} A^+_{i}A^-_{j} + \frac{g^2}{4} (A_i^+ A^{-i})^2 - \frac{g^2}{4} ( A_i^+ A^{+i} ) (A_i^- A^{- i})
\end{multline}
where we defined
\begin{align}
    G_{ij}^+ &= (G_{ij}^-)^\dagger = A_{ij}^+ + ig  (A^+_{i}A^3_{j} - A^+_{j}A^3_{i})   \\
    A_{ij}^+ &= (A_{ij}^-)^\dagger = \partial_{i}A^+_{j} - \partial_{j}A^+_{i} \\
    Z_{ij} &= A^3_{ij} = \partial_{i}A^3_{j} - \partial_{j}A^3_{i}
\end{align}
Looking at the terms generating a mass term for the $A^\pm$ boson, these amount to:
\begin{equation}
    \left( \frac{g^2}{2} |{h}_1|^2 \delta^{ij} + ig Z^{ij} + g^2 (\phi^3)^2 \right) A_i^+ A_j^-
\end{equation}
or after diagonalisation, the effective mass term is given by:
\begin{align}
    m_{W \text{eff}}^2 &= m_W^2 + m_Z^2 \pm gB_Z 
\end{align}
where $B_Z =\sqrt{Z_{ij}Z^{ij}} \approx  \Phi_B / (\pi R_S^2)$ via Eq. \eqref{Magnetic flux of cosmic string} for $R_S$ the string width. Since $R_S \sim m_Z^{-1}$, we see therefore that lowering the hierarchy $\eta$ amounts to an increased instability due to the fact that the $A^\pm$ bosons acquire a negative mass in the core of the string. Note that the mass has two branches corresponding to the spin alignment of the vector with the string magnetic field.
This explains the $A^\pm$ driven instability: as the mass hierarchy $\eta$ decreases, the effective mass of the $A^\pm$ becomes smaller, and the excitation modes in the $A^\pm$ channel become more accessible. This instability can be traced back through the stability equations via the term $-4n\zeta'(\rho)/\rho$ appearing in \eqref{D4 stability eqn} which yields a negative contribution to the eigenvalue in the vicinity of the core.

Once a certain threshold attained, the string core vacuum becomes a tachyon direction, and a $A^\pm$-condensate forms,
analogous to the 
$W$-condensation instability of the electroweak string
(we briefly review the instability of the electroweak string in Appendix \ref{App electroweak string}) \cite{Ambjorn:1989bd, Garriga:1995fv}. This explains why the instability region is increased for smaller $\eta$ values.





\subsection{Comparison with literature}\label{Appendix comparison Chitose}
\begin{figure}[h!]
    \centering
    \includegraphics[width=10cm]{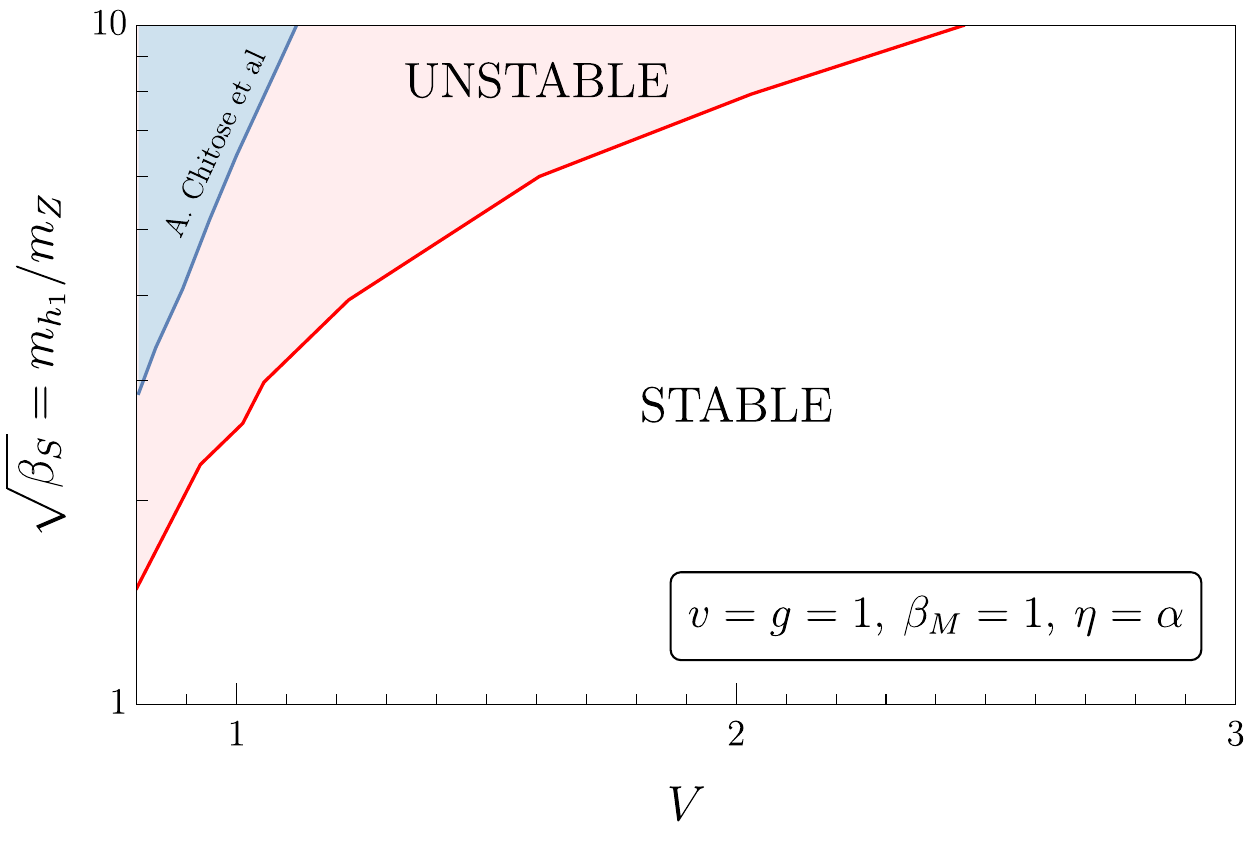}
    \caption{Comparison of our predictions of the stability of the cosmic string with the classical stability presented in \cite{Chitose:2023dam} in the case for which $g=v=1$ and $m_{\phi^3} = m_{h_2} = m_W$. Our analysis predicts a larger region of classical instability.}
    \label{Fig LSA Comparison with Chitose}
\end{figure}

Finally, we compare the results of our stability analysis with a benchmark previously studied in \cite{Chitose:2023dam}, where a specific unwinding ansatz was considered following \cite{Shifman:2002yi}. The benchmark corresponds to taking $g=1, V=1$ and $m_{\phi^3} = m_{h_2} = m_{W}$, or in other words: $\alpha=\eta$ and $\beta_M=1$. We compare our predictions and that of the literature in Figure \ref{Fig LSA Comparison with Chitose}. We observe that our results predict a larger region of classical instability.

\section{Instability of the electroweak string}\label{App electroweak string}
The electroweak sector of the Standard Model also allows for cosmic string solutions, albeit non-topological ones. We recall that the Lagrangian of this gauge theory $SU(2)\times U(1)$ is
\begin{equation}
    \mathcal{L} = (D_\mu h)^\dagger D^\mu h - \frac{1}{4}W_{\mu\nu}^a W^{a\mu\nu} - \frac{1}{4}B_{\mu \nu} B^{\mu\nu} - V(h),
\end{equation}
where the covariant derivative and the field strengths can be listed as
\begin{align}
    D_\mu h &= \left( \partial_\mu - ig \frac{\sigma^a}{2}W_\mu^a - i g' \frac{Y}{2}B_\mu \right)h, \\
    W_{\mu\nu}^a &= \partial_\mu W_\nu^a - \partial_\nu W_\mu^a + g \epsilon^{abc} W_{\mu}^b W_{\nu}^c, \\
    B_{\mu\nu} &= \partial_\mu B_\nu - \partial_\nu B_\mu,
\end{align}
and with the potential containing a single Mexican hat for the Higgs doublet $h$:
\begin{equation}
    V(h) = \lambda \left(h^\dagger h - v² \right)^2
\end{equation} 
The physical fields ($W^\pm, Z$ and $\gamma$) are identified by the linear combinations given by
\begin{align}
    W_\mu^\pm &= (W_\mu^1 \mp iW_\mu^2)/\sqrt{2}, \\
    Z_\mu &= \cos \theta_W \, W_\mu^3 - \sin\theta_W \, B_\mu, \\
    A_\mu &= \sin\theta_W \, W_\mu^3 + \cos\theta_W \, B_\mu,
\end{align}
respectively. Finally, the spontaneous symmetry breaking makes that the particles acquire masses given by: $m_H = \sqrt{2\lambda} v$, $m_W = gv/2$ and $m_Z = g_Z v/2$ for $g_Z = \sqrt{g^2 + g'^2}$. The weak mixing angle is related to the gauge couplings via $\sin\theta_W = g'/g_Z$.

Similarly to Eq. \eqref{Standard string ansatz}, the electroweak string is obtained by making the following ansatz on the fields:
\begin{equation}
    h = vf(\rho) e^{in\theta} \begin{pmatrix}
        0 \\ 1
    \end{pmatrix}, \quad Z_\mu = \frac{n}{g\rho}\zeta(\rho)\delta_{\mu\theta}, \quad A_\mu = W_\mu^\pm = 0.
    \label{EW String ansatz}
\end{equation}
Note however that we now selected the lower component of $h$ to form the vortex geometry in order to use the $Z-$boson to screen its otherwise infinite gradient energy at infinity.

\begin{figure}[t!]
    \centering
    \includegraphics[width=12cm]{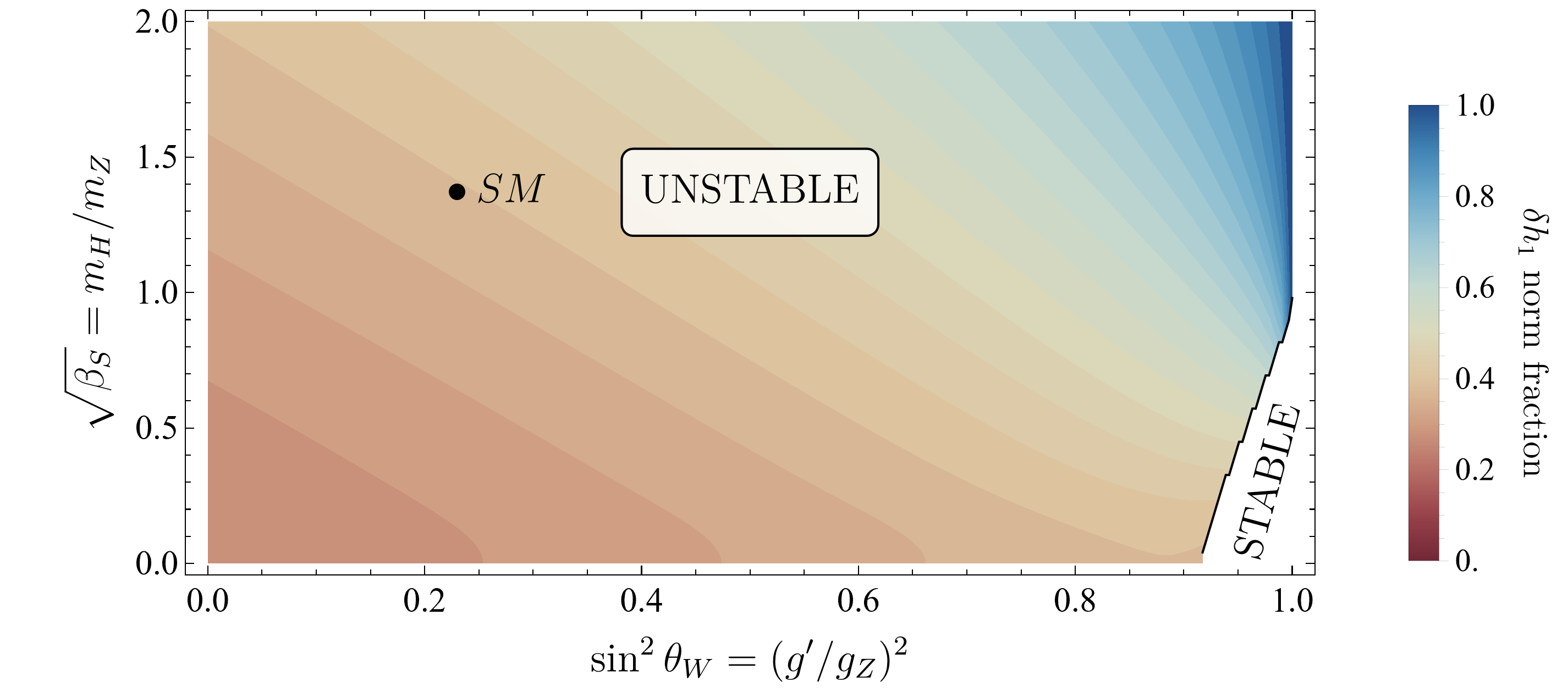}
    \caption{Classical stability analysis of the electroweak string in the parameter space of the model with $\sin^2\theta_W = (g'/g_Z)^2$ and $\sqrt{\beta_S} = m_H/m_Z$. The white region in the corner indicates when the string ansatz of Eq. \eqref{EW String ansatz} is classically stable. For the Standard Model (SM), the electroweak string lies deep in the unstable region. The color gradient indicates the norm fraction $\braket{a_{l=0}, a_{l=0}}/\braket{\delta \Psi_{l=0}, \delta \Psi_{l=0}}$ in the unstable regions. Bluer regions indicate a norm domination of the $\delta h_1$ perturbation via the $a_{l=0}$ profile, while red regions highlight regions dominated by $\delta W_\downarrow$ perturbations.
    }
    \label{Fig EW String LSA}
\end{figure}

In straight analogy to the analysis above, the electroweak string is non-topological, and its classical stability can only be ensured dynamically. Therefore, one needs to investigate its classical stability explicitly. We refer to \cite{Goodband:1995he, James:1992wb} for an extensive review of the stability analysis. The same analysis as outlined in Section \ref{Section stability} was performed for the electroweak string for which we have to solve an eigenvalue equation of the type
\begin{equation}
    \mathcal{D}_{\text{EW}} \begin{pmatrix}
        \delta h_1 \\
        \delta W_\mu^{+} 
    \end{pmatrix} = \omega^2 \begin{pmatrix}
        \delta h_1 \\
        \delta W_\mu^{+} 
    \end{pmatrix}
\end{equation}
Our results are shown in Figure \ref{Fig EW String LSA} as a function of the dynamical ratio $\beta_S \equiv (m_H/m_Z)² = 8\lambda / g_Z^2$ and the weak mixing angle via $\sin^2\theta_W$. The only stable region is the bottom right corner for values of $\beta<1$ and in the limit of $\sin^2\theta_W >0.9$. In particular, we retrieve the same stability region for the semi-local strings (charged under $SU(2)_{\text{global}}\times U(1)_\text{gauged}$) in the limit of $\sin^2\theta_W = 1$. The parameter space is clearly dominated by instabilities in the electroweak string. In addition to this, to have more insights we indicated the L2-norm on the obtained eigenvalues in case of an instability, similarly as what we did previously in Figure \ref{Fig L2 Domination Plot}. We observe that the instabilities are largely dominated by perturbations from the $W$-boson. In particular, for the parameters accommodating the values of the Standard Model of particle physics ($m_H \approx 125$ GeV and $m_Z \approx 91$ GeV), the electroweak string is highly unstable, and the instability is driven through the condensation of the $W$-boson in the core. We expect this decay channel to be analogous to the one we observe in the model of metastable strings in the limit of a small, reversed hierarchy $V<v$, i.e. $\eta <2$: see Appendix \ref{Appendix low eta limit}.

\section{Stability analysis and time evolution in the global theory}\label{Appendix Global Model}
We now consider the global variant of the gauged theory given in Eq. \eqref{Lagrangian full theory OWN model} by putting the gauge fields to zero and only consider the scalars as the dynamical degrees of freedom. This will allow us to study the time evolution of the perturbed string in a simpler model. The
Lagrangian of the global theory is 
\begin{equation}
    \mathcal{L} = \frac{1}{2}\partial_\mu \phi^a \partial^\mu \phi^a + \partial_\mu h^\dagger \partial^\mu h - V(\phi, h),
\end{equation}
where the potential $V(\phi, h)$ is still given by Eq. \eqref{Potential full theory own model}. The equations of motion are reduced to the following set of five coupled differential equations
\begin{align}
    \partial_{\mu}\partial^{\mu}\phi^{a} &= -4\tilde{\lambda}\left(\phi^{b}\phi^{b}-V^2\right)\phi^{a} - \frac{\gamma}{2}h^{\dagger}\left(\phi^{a}-V\sigma^{a}\right)h \label{EOM Global model in h},\\
    \partial_{\mu}\partial^{\mu}h &= -2\lambda\left(h^\dagger h -v^2\right)h -\frac{\gamma}{4}\left(\phi^a \sigma^a - V\right)^2 h
\end{align}
The string solution we now study is the same as in Eq. \eqref{Standard string ansatz} but without the presence of the gauge fields. We therefore already expect a logarithmically diverging tension due to the gradient energy of $h_1$.

\paragraph{The linear stability}
Perturbing the equations of motion above results in a decoupling of various sectors. In particular, the sector that gives instabilities related to the portal term has as stability equations:
\begin{align}
    \left( \Box + 2\lambda(|\bar{h}_1|^2 - v^2) \right)\delta h_2 - \frac{\gamma V}{\sqrt{2}}\bar{h}_1^\dagger \delta h_2 = 0 \\
    \left( \Box + \frac{\gamma}{2}|\bar{h}_1|^2 \right)\delta\phi^- - \frac{\gamma V}{\sqrt{2}}\delta h_2 = 0
\end{align}
In the global theory the equations are rescaled differently according to 
\begin{equation}
    x\to \tilde{x} = m_{h_1}x, \qquad \Psi \to \tilde{\Psi} = \Psi/v.
\end{equation}
To make further progress, we substitute the following ansatz for the perturbations which is but the same ansatz as we used in the full model (see Eqs. \eqref{Ansatz 1 perturbation full model} and \eqref{Ansatz 2 perturbation full model}):
\begin{align}
    \delta h_2 &= s(\rho) e^{i l \theta} e^{i \omega t},\\
    \delta \phi^- &= a(\rho) e^{i (l-n) \theta} e^{i \omega t}.
\end{align}
This yields the following coupled eigenvalue differential equation using the above rescaling:
\begin{equation}
\label{eq:eigen_big}
    \begin{pmatrix}
        D_{1} & A\\
        A & D_{2}
    \end{pmatrix} \begin{pmatrix}
        s \\ a
    \end{pmatrix} = \omega^2 \begin{pmatrix}
        s \\ a
    \end{pmatrix}
\end{equation}
where the elements of this stability matrix are given by:
\begin{align}
    D_{1} &= -\frac{d^2}{d\rho^2} - \frac{1}{\rho} \frac{d}{d\rho} + \frac{l^2}{\rho^2} + \frac{1}{2}\left(f^2(\rho)-1\right) + \xi_1\xi_2 \\
    D_{2} &= -\frac{d^2}{d\rho^2} - \frac{1}{\rho}\frac{d}{d\rho} + \frac{(l-n)^2}{\rho^2} + \xi_1 f^2(\rho) \\
    A &= -\xi_1 \sqrt{\xi_2} f(\rho)
\end{align}
The numerical results of this stability analysis were outlined in Figure \ref{fig:lsa_global_model}.

\paragraph{Time evolution of the perturbed string}
In order to study the late-time outcome of the perturbations (i.e. stationary condensate or full dissipation), we solve for the equations of motion using a staggered leapfrog method. Using the above rescaling, the equations of motion are written explicitly via
\begin{align}
\ddot{h}_1 &=
\begin{aligned}[t]
\nabla^2 h_1 -D\dot{h}_1
&- \frac{1}{2}(h^\dagger h - 1)\, h_1 \\
&- \frac{\xi_1}{2} \Big[
\big((\phi^1)^2 + (\phi^2)^2 + (\phi^3 - \sqrt{\eta/2})^2\big)\, h_1  - \sqrt{2 \eta}\, (\phi^1 - i \phi^2)\, h_2
\Big]
\end{aligned}
\\[0.5em]
\ddot{h}_2 &=
\begin{aligned}[t]
\nabla^2 h_2 - D&\dot{h}_2
- \frac{1}{2}(h^\dagger h - 1)\, h_2 \\
&- \frac{\xi_1}{2} \Big[
- \sqrt{2 \eta}\, (\phi^1 + i \phi^2)\, h_1  + \big((\phi^1)^2 + (\phi^2)^2 + (\phi^3 + \sqrt{\eta/2})^2\big)\, h_2
\Big]
\end{aligned}
\\[0.5em]
\ddot{\phi}^1 &=
\begin{aligned}[t]
\nabla^2 \phi^1 -D\dot{\phi}^1 &- \xi_3 (\phi^a \phi^a - \xi_2/2)\, \phi^1 \\
&- \xi_1 \Big[
\phi^1 (|h_1|^2 + |h_2|^2) - \sqrt{\xi_2/2}\, (h_1^\dagger h_2 + h_2^\dagger h_1)
\Big]
\end{aligned}
\\[0.5em]
\ddot{\phi}^2 &=
\begin{aligned}[t]
\nabla^2 \phi^2 -D\dot{\phi}^2 &- \xi_3 (\phi^a \phi^a - \xi_2/2)\, \phi^2 \\
&- \xi_1 \Big[
\phi^2 (|h_1|^2 + |h_2|^2) + i \sqrt{\xi_2/2}\, (h_1^\dagger h_2 - h_2^\dagger h_1)
\Big]
\end{aligned}
\\[0.5em]
\ddot{\phi}^3 &=
\begin{aligned}[t]
\nabla^2 \phi^3 -D\dot{\phi}^3 &- \xi_3 (\phi^a \phi^a - \xi_2/2)\, \phi^3 \\
&- \xi_1 \Big[
(\phi^3 - \sqrt{\xi_2/2})\, |h_1|^2 + (\phi^3 + \sqrt{\xi_2/2})\, |h_2|^2
\Big]
\end{aligned}
\end{align}
where we defined the following set of three dimensionless fundamental parameters
\begin{equation}
    \xi_1 = \frac{\gamma}{8\lambda} = \frac{\alpha}{\beta_S \eta}, \qquad \xi_2 = \frac{2V^2}{v^2} = \eta, \qquad \xi_3 = \frac{\tilde{\lambda}}{\lambda} = \frac{\beta_M}{\beta_S}.
\end{equation}
Note that we also inserted friction terms in order to suppress local spurious oscillations and speed up the physical time evolution. In our computations, we systematically take the friction coefficient to be $D=0.1$. Lower values allow for oscillations due to the leapfrog algorithm to be omnipresent in the profiles, while larger values make the dynamics slower without changing the overall physics.  In the discretization procedure, the friction terms are implemented via a Crank-Nicolson method: see \cite{Figueroa:2020rrl} which we follow in this matter.

Lastly, the time evolutions for the profiles we obtained in Section \ref{Section chasing the instability} were obtained by considering the perturbed string as initial condition:
\begin{equation}
    \Psi(x,y) = \bar{\Psi}(x,y) + \epsilon \left.\delta \Psi\right|_{\text{min } \omega2}(x,y),
\end{equation}
where we take $\epsilon=0.01$ and $\left.\delta \Psi\right|_{\text{min } \omega2}$ to be the normalized eigenfunction corresponding to the lowest eigenvalue (the negative eigenmode in case of an unstable string). The initial field velocity is taken to be zero. The initial profiles are interpolated on a cartesian 2D grid and subsequently evolved without constraining the core. At every time step, we enforce Neumann boundary conditions on the fields: $\partial_n\Psi_{\text{boundary}} = 0$ for $n$ the normal direction.

\paragraph{The (rescaled) energy density}
In order to test our numerical results, we need to test conservation of energy. The total energy, where the integrand (i.e. the energy density) is written in rescaled units is given by:
\begin{align}
    E = v^2 \int dx dy\left[ |\dot{h}|^2 + \tfrac{1}{2}(\dot{\phi}^a)^2 + \tfrac{1}{2}(\partial_i \phi^a)(\partial_i \phi^a) + (\partial_i h)^\dagger (\partial_i h)\
    \right.\\
    \left. + \tfrac{1}{4}(h^\dagger h-1)^2 + \tfrac{1}{4}\xi_3 (\phi^a \phi^a - \xi_2/2)^2 + \tfrac{1}{2}\xi_1 h^\dagger\left(\phi^a\sigma^a -\sqrt{\xi_2/2}\right)^2 h\right] 
\end{align}
Note that the kinetic terms from the extra field velocity terms are not included in the definition of the string tension used for Figure \ref{fig:tension_evolution}. In the case of $D=0$ we observe that the total energy oscillates around the initial value. When $D=0.1$, the energy monotonically decreases, as expected.

\end{appendices}

\bibliographystyle{JHEP}
\bibliography{sample.bib}

@article{LIGOScientific:2025kry,
    author = "Abac, A. G. and others",
    collaboration = "LIGO Scientific, VIRGO, KAGRA",
    title = "{Cosmological and High Energy Physics implications from gravitational-wave background searches in LIGO-Virgo-KAGRA's O1-O4a runs}",
    eprint = "2510.26848",
    archivePrefix = "arXiv",
    primaryClass = "gr-qc",
    reportNumber = "LIGO-PP2500150",
    month = "10",
    year = "2025"
}

@article{Abrikosov:1956sx,
    author = "Abrikosov, A. A.",
    title = "{On the Magnetic Properties of Superconductors of the Second Group}",
    journal = "Sov. Phys. JETP",
    volume = "5",
    pages = "1174--1182",
    year = "1957"
}

@article{Figueroa:2020rrl,
    author = "Figueroa, Daniel G. and Florio, Adrien and Torrenti, Francisco and Valkenburg, Wessel",
    title = "{The art of simulating the early Universe -- Part I: Integration techniques and canonical cases}",
    eprint = "2006.15122",
    archivePrefix = "arXiv",
    primaryClass = "astro-ph.CO",
    doi = "10.1088/1475-7516/2021/04/035",
    journal = "JCAP",
    volume = "04",
    pages = "035",
    year = "2021"
}

@article{King:2021gmj,
    author = "King, Stephen F. and Pascoli, Silvia and Turner, Jessica and Zhou, Ye-Ling",
    title = "{Confronting SO(10) GUTs with proton decay and gravitational waves}",
    eprint = "2106.15634",
    archivePrefix = "arXiv",
    primaryClass = "hep-ph",
    reportNumber = "IPPP/20/120",
    doi = "10.1007/JHEP10(2021)225",
    journal = "JHEP",
    volume = "10",
    pages = "225",
    year = "2021"
}

@article{King:2020hyd,
    author = "King, Stephen F. and Pascoli, Silvia and Turner, Jessica and Zhou, Ye-Ling",
    title = "{Gravitational Waves and Proton Decay: Complementary Windows into Grand Unified Theories}",
    eprint = "2005.13549",
    archivePrefix = "arXiv",
    primaryClass = "hep-ph",
    reportNumber = "FERMILAB-PUB-20-187-T, IPPP/20/20",
    doi = "10.1103/PhysRevLett.126.021802",
    journal = "Phys. Rev. Lett.",
    volume = "126",
    number = "2",
    pages = "021802",
    year = "2021"
}

@article{Fu:2023mdu,
    author = "Fu, Bowen and King, Stephen F. and Marsili, Luca and Pascoli, Silvia and Turner, Jessica and Zhou, Ye-Ling",
    title = "{Testing realistic SO(10) SUSY GUTs with proton decay and gravitational waves}",
    eprint = "2308.05799",
    archivePrefix = "arXiv",
    primaryClass = "hep-ph",
    reportNumber = "IPPP/23/41",
    doi = "10.1103/PhysRevD.109.055025",
    journal = "Phys. Rev. D",
    volume = "109",
    number = "5",
    pages = "055025",
    year = "2024"
}

@article{Ingoldby:2025wcl,
    author = "Ingoldby, James and Khoze, Valentin V. and Turner, Jessica",
    title = "{Metastable Strings and Gravitational Waves in One-Scale Models}",
    eprint = "2511.08546",
    archivePrefix = "arXiv",
    primaryClass = "hep-ph",
    reportNumber = "IPPP/25/68",
    month = "11",
    year = "2025"
}

@article{Dunsky:2021tih,
    author = "Dunsky, David I. and Ghoshal, Anish and Murayama, Hitoshi and Sakakihara, Yuki and White, Graham",
    title = "{GUTs, hybrid topological defects, and gravitational waves}",
    eprint = "2111.08750",
    archivePrefix = "arXiv",
    primaryClass = "hep-ph",
    doi = "10.1103/PhysRevD.106.075030",
    journal = "Phys. Rev. D",
    volume = "106",
    number = "7",
    pages = "075030",
    year = "2022"
}

@article{ET:2025xjr,
    author = "Abac, Adrian and others",
    collaboration = "ET",
    title = "{The Science of the Einstein Telescope}",
    eprint = "2503.12263",
    archivePrefix = "arXiv",
    primaryClass = "gr-qc",
    reportNumber = "ET-0036C-25",
    month = "3",
    year = "2025"
}

@article{Blanco-Pillado:2024aca,
    author = "Blanco-Pillado, Jose J. and Cui, Yanou and Kuroyanagi, Sachiko and Lewicki, Marek and Nardini, Germano and Pieroni, Mauro and Rybak, Ivan Yu. and Sousa, Lara and Wachter, Jeremy M.",
    collaboration = "LISA Cosmology Working Group",
    title = "{Gravitational waves from cosmic strings in LISA: reconstruction pipeline and physics interpretation}",
    eprint = "2405.03740",
    archivePrefix = "arXiv",
    primaryClass = "astro-ph.CO",
    reportNumber = "LISA-COSWG-24-02, CERN-TH-2024-085",
    doi = "10.1088/1475-7516/2025/05/006",
    journal = "JCAP",
    volume = "05",
    pages = "006",
    year = "2025"
}

@article{NANOGrav:2023hvm,
    author = "Afzal, Adeela and others",
    collaboration = "NANOGrav",
    title = "{The NANOGrav 15 yr Data Set: Search for Signals from New Physics}",
    eprint = "2306.16219",
    archivePrefix = "arXiv",
    primaryClass = "astro-ph.HE",
    reportNumber = "FERMILAB-PUB-23-589-T",
    doi = "10.3847/2041-8213/acdc91",
    journal = "Astrophys. J. Lett.",
    volume = "951",
    number = "1",
    pages = "L11",
    year = "2023",
    note = "[Erratum: Astrophys.J.Lett. 971, L27 (2024), Erratum: Astrophys.J. 971, L27 (2024)]"
}

@article{Zurek:1985qw,
    author = "Zurek, W. H.",
    title = "{Cosmological Experiments in Superfluid Helium?}",
    doi = "10.1038/317505a0",
    journal = "Nature",
    volume = "317",
    pages = "505--508",
    year = "1985"
}

@article{Zeldovich:1974uw,
    author = "Zeldovich, Ya. B. and Kobzarev, I. Yu. and Okun, L. B.",
    title = "{Cosmological Consequences of the Spontaneous Breakdown of Discrete Symmetry}",
    reportNumber = "SLAC-TRANS-0165, IPM-MOSCOW-15",
    journal = "Zh. Eksp. Teor. Fiz.",
    volume = "67",
    pages = "3--11",
    year = "1974"
}

@article{Chitose:2023dam,
    author = "Chitose, Akifumi and Ibe, Masahiro and Nakayama, Yuhei and Shirai, Satoshi and Watanabe, Keiichi",
    title = "{Revisiting metastable cosmic string breaking}",
    eprint = "2312.15662",
    archivePrefix = "arXiv",
    primaryClass = "hep-ph",
    reportNumber = "IPMU23-0054",
    doi = "10.1007/JHEP04(2024)068",
    journal = "JHEP",
    volume = "04",
    pages = "068",
    year = "2024"
}

@article{Achucarro:1999it,
    author = "Achucarro, Ana and Vachaspati, Tanmay",
    title = "{Semilocal and electroweak strings}",
    eprint = "hep-ph/9904229",
    archivePrefix = "arXiv",
    reportNumber = "EHU-FT-9808A, CWRU-P34-1998",
    doi = "10.1016/S0370-1573(99)00103-9",
    journal = "Phys. Rept.",
    volume = "327",
    pages = "347--426",
    year = "2000"
}

@article{James:1992zp,
    author = "James, M. and Perivolaropoulos, L. and Vachaspati, T.",
    title = "{Stability of electroweak strings}",
    doi = "10.1103/PhysRevD.46.R5232",
    journal = "Phys. Rev. D",
    volume = "46",
    pages = "R5232--R5235",
    year = "1992"
}

@article{James:1992wb,
    author = "James, Margaret and Perivolaropoulos, Leandros and Vachaspati, Tanmay",
    title = "{Detailed stability analysis of electroweak strings}",
    eprint = "hep-ph/9212301",
    archivePrefix = "arXiv",
    reportNumber = "CFA-3542",
    doi = "10.1016/0550-3213(93)90046-R",
    journal = "Nucl. Phys. B",
    volume = "395",
    pages = "534--546",
    year = "1993"
}

@article{Earnshaw:1993yu,
    author = "Earnshaw, Michael A. and James, Margaret",
    title = "{Stability of two doublet electroweak strings}",
    eprint = "hep-ph/9308223",
    archivePrefix = "arXiv",
    reportNumber = "DAMTP-93-39",
    doi = "10.1103/PhysRevD.48.5818",
    journal = "Phys. Rev. D",
    volume = "48",
    pages = "5818--5826",
    year = "1993"
}

@article{Masperi:1993fw,
    author = "Masperi, Luis and Megevand, Ariel and Savaglio, Sandra",
    title = "{Stability of modified electroweak strings}",
    eprint = "hep-ph/9410211",
    archivePrefix = "arXiv",
    reportNumber = "PRINT-94-0016 (BARILOCHE)",
    doi = "10.1007/s002880050234",
    journal = "Z. Phys. C",
    volume = "72",
    pages = "171--174",
    year = "1996"
}

@article{Forgacs:2019tbn,
    author = "Forg{\'a}cs, P{\'e}ter and Luk{\'a}cs, {\'A}rp{\'a}d",
    title = "{Electroweak strings with dark scalar condensates and their stability}",
    eprint = "1909.07447",
    archivePrefix = "arXiv",
    primaryClass = "hep-ph",
    doi = "10.1103/PhysRevD.102.023009",
    journal = "Phys. Rev. D",
    volume = "102",
    number = "2",
    pages = "023009",
    year = "2020"
}

@article{Hindmarsh:1991jq,
    author = "Hindmarsh, Mark",
    title = "{Existence and stability of semilocal strings}",
    reportNumber = "NCL-91-TP7",
    doi = "10.1103/PhysRevLett.68.1263",
    journal = "Phys. Rev. Lett.",
    volume = "68",
    pages = "1263--1266",
    year = "1992"
}

@article{Vachaspati:1992jk,
    author = "Vachaspati, Tanmay",
    title = "{Electroweak strings}",
    reportNumber = "TUTP-92-3",
    doi = "10.1016/0550-3213(93)90189-V",
    journal = "Nucl. Phys. B",
    volume = "397",
    pages = "648--671",
    year = "1993"
}

@article{Goodband:1995he,
    author = "Goodband, Michael and Hindmarsh, Mark",
    title = "{Instabilities of electroweak strings}",
    eprint = "hep-ph/9505357",
    archivePrefix = "arXiv",
    reportNumber = "SUSX-TH-95-72",
    doi = "10.1016/0370-2693(95)01198-Y",
    journal = "Phys. Lett. B",
    volume = "363",
    pages = "58--64",
    year = "1995"
}

@article{Vachaspati:1991dz,
    author = "Vachaspati, T. and Achucarro, A.",
    title = "{Semilocal cosmic strings}",
    doi = "10.1103/PhysRevD.44.3067",
    journal = "Phys. Rev. D",
    volume = "44",
    pages = "3067--3071",
    year = "1991"
}

@article{Achucarro:1993bu,
    author = "Achucarro, Ana and Gregory, Ruth and Harvey, Jeffrey A. and Kuijken, Konrad",
    title = "{Cinderella strings}",
    eprint = "hep-th/9312034",
    archivePrefix = "arXiv",
    reportNumber = "EFI-93-68",
    doi = "10.1103/PhysRevLett.72.3646",
    journal = "Phys. Rev. Lett.",
    volume = "72",
    pages = "3646--3649",
    year = "1994"
}

@article{Garaud:2010ng,
    author = "Garaud, Julien and Volkov, Mikhail S.",
    title = "{Stability Analysis of Superconducting Electroweak Vortices}",
    eprint = "1005.3002",
    archivePrefix = "arXiv",
    primaryClass = "hep-th",
    doi = "10.1016/j.nuclphysb.2010.06.016",
    journal = "Nucl. Phys. B",
    volume = "839",
    pages = "310--340",
    year = "2010"
}

@article{Shifman:2002yi,
    author = "Shifman, M. and Yung, Alexei",
    title = "{Metastable strings in Abelian Higgs models embedded in nonAbelian theories: Calculating the decay rate}",
    eprint = "hep-th/0205025",
    archivePrefix = "arXiv",
    reportNumber = "TPI-MINN-02-11, UMN-TH-2051-02",
    doi = "10.1103/PhysRevD.66.045012",
    journal = "Phys. Rev. D",
    volume = "66",
    pages = "045012",
    year = "2002"
}

@article{Eto:2024xvc,
    author = "Eto, Minoru and Hamada, Yu and Jinno, Ryusuke and Nitta, Muneto and Yamada, Masatoshi",
    title = "{Neutrino zeromodes on electroweak strings in light of topological insulators}",
    eprint = "2402.19417",
    archivePrefix = "arXiv",
    primaryClass = "hep-ph",
    reportNumber = "KEK-TH-2592, DESY-24-003, RESCEU-2/24, YGHP-24-01",
    doi = "10.1007/JHEP06(2024)062",
    journal = "JHEP",
    volume = "06",
    pages = "062",
    year = "2024"
}

@article{Polyakov:1974ek,
    author = "Polyakov, Alexander M.",
    editor = "Taylor, J. C.",
    title = "{Particle Spectrum in Quantum Field Theory}",
    reportNumber = "PRINT-74-1566 (LANDAU-INST)",
    journal = "JETP Lett.",
    volume = "20",
    pages = "194--195",
    year = "1974"
}

@book{Shnir:2005vvi,
    author = "Shnir, Yakov M.",
    title = "{Magnetic Monopoles}",
    doi = "10.1007/3-540-29082-6",
    isbn = "978-3-540-25277-1, 978-3-540-29082-7",
    publisher = "Springer",
    address = "Berlin/Heidelberg",
    series = "Text and Monographs in Physics",
    year = "2005"
}

@article{Goodband:1995rt,
    author = "Goodband, Michael and Hindmarsh, Mark",
    title = "{Bound states and instabilities of vortices}",
    eprint = "hep-ph/9503457",
    archivePrefix = "arXiv",
    reportNumber = "SUSX-TH-95-71",
    doi = "10.1103/PhysRevD.52.4621",
    journal = "Phys. Rev. D",
    volume = "52",
    pages = "4621--4632",
    year = "1995"
}

@book{Shifman:2012zz,
    author = "Shifman, Mikhail",
    title = "{Advanced topics in quantum field theory.}: {A lecture course}",
    doi = "10.1017/9781108885911",
    isbn = "978-1-139-21036-2, 978-0-521-19084-8, 978-1-108-88591-1, 978-1-108-84042-2",
    publisher = "Cambridge Univ. Press",
    address = "Cambridge, UK",
    month = "2",
    year = "2012"
}

@article{Kirkman:1981ck,
    author = "Kirkman, Thomas W. and Zachos, Cosmas K.",
    title = "{Asymptotic Analysis of the Monopole Structure}",
    reportNumber = "DOE-ER/00881-207",
    doi = "10.1103/PhysRevD.24.999",
    journal = "Phys. Rev. D",
    volume = "24",
    pages = "999",
    year = "1981"
}

@article{Derrick:1964ww,
    author = "Derrick, G. H.",
    title = "{Comments on nonlinear wave equations as models for elementary particles}",
    doi = "10.1063/1.1704233",
    journal = "J. Math. Phys.",
    volume = "5",
    pages = "1252--1254",
    year = "1964"
}

@article{Eto:2016mqc,
    author = "Eto, Minoru and Nitta, Muneto and Sakurai, Kohei",
    title = "{Stabilizing semilocal strings by polarization}",
    eprint = "1608.03516",
    archivePrefix = "arXiv",
    primaryClass = "hep-th",
    reportNumber = "YGHP-16-05",
    doi = "10.1007/JHEP10(2016)048",
    journal = "JHEP",
    volume = "10",
    pages = "048",
    year = "2016"
}

@article{Forgacs:2016dby,
    author = "Forg{\'a}cs, P{\'e}ter and Luk{\'a}cs, {\'A}rp{\'a}d",
    title = "{Stabilization of semilocal strings by dark scalar condensates}",
    eprint = "1612.03151",
    archivePrefix = "arXiv",
    primaryClass = "hep-th",
    doi = "10.1103/PhysRevD.95.035003",
    journal = "Phys. Rev. D",
    volume = "95",
    number = "3",
    pages = "035003",
    year = "2017"
}

@article{Abe:2020ure,
    author = "Abe, Yoshihiko and Hamada, Yu and Yoshioka, Koichi",
    title = "{Electroweak axion string and superconductivity}",
    eprint = "2010.02834",
    archivePrefix = "arXiv",
    primaryClass = "hep-ph",
    reportNumber = "KUNS-2838",
    doi = "10.1007/JHEP06(2021)172",
    journal = "JHEP",
    volume = "06",
    pages = "172",
    year = "2021"
}

@article{Eto:2021dca,
    author = "Eto, Minoru and Hamada, Yu and Nitta, Muneto",
    title = "{Stable Z-strings with topological polarization in two Higgs doublet model}",
    eprint = "2111.13345",
    archivePrefix = "arXiv",
    primaryClass = "hep-ph",
    reportNumber = "YGHP-21-4, KEK-TH-2370",
    doi = "10.1007/JHEP02(2022)099",
    journal = "JHEP",
    volume = "02",
    pages = "099",
    year = "2022"
}

@article{Kanda:2023yyz,
    author = "Kanda, Yukihiro and Maekawa, Nobuhiro",
    title = "{Stability of the embedded string in the SU(N){\texttimes}U(1) Higgs model and its application}",
    eprint = "2303.09517",
    archivePrefix = "arXiv",
    primaryClass = "hep-ph",
    doi = "10.1103/PhysRevD.107.096007",
    journal = "Phys. Rev. D",
    volume = "107",
    number = "9",
    pages = "096007",
    year = "2023"
}

@article{Garriga:1995fv,
    author = "Garriga, Jaume and Montes, Xavi",
    title = "{Stability of Z strings in strong magnetic fields}",
    eprint = "hep-ph/9505424",
    archivePrefix = "arXiv",
    reportNumber = "UAB-FT-367",
    doi = "10.1103/PhysRevLett.75.2268",
    journal = "Phys. Rev. Lett.",
    volume = "75",
    pages = "2268--2271",
    year = "1995"
}

@article{Kibble:1976sj,
    author = "Kibble, T. W. B.",
    title = "{Topology of Cosmic Domains and Strings}",
    reportNumber = "ICTP/75/5",
    doi = "10.1088/0305-4470/9/8/029",
    journal = "J. Phys. A",
    volume = "9",
    pages = "1387--1398",
    year = "1976"
}

@article{Preskill:1992ck,
    author = "Preskill, John and Vilenkin, Alexander",
    title = "{Decay of metastable topological defects}",
    eprint = "hep-ph/9209210",
    archivePrefix = "arXiv",
    reportNumber = "HUTP-92-A018, CALT-68-1786",
    doi = "10.1103/PhysRevD.47.2324",
    journal = "Phys. Rev. D",
    volume = "47",
    pages = "2324--2342",
    year = "1993"
}

@book{Vilenkin:2000jqa,
    author = "Vilenkin, A. and Shellard, E. P. S.",
    title = "{Cosmic Strings and Other Topological Defects}",
    isbn = "978-0-521-65476-0",
    publisher = "Cambridge University Press",
    month = "7",
    year = "2000"
}

@article{Buchmuller:2019gfy,
    author = "Buchmuller, Wilfried and Domcke, Valerie and Murayama, Hitoshi and Schmitz, Kai",
    title = "{Probing the scale of grand unification with gravitational waves}",
    eprint = "1912.03695",
    archivePrefix = "arXiv",
    primaryClass = "hep-ph",
    reportNumber = "CERN-TH-2019-215, DESY-19-210, DESY 19-210, IPMU 19-0179",
    doi = "10.1016/j.physletb.2020.135764",
    journal = "Phys. Lett. B",
    volume = "809",
    pages = "135764",
    year = "2020"
}

@article{Buchmuller:2023aus,
    author = "Buchmuller, Wilfried and Domcke, Valerie and Schmitz, Kai",
    title = "{Metastable cosmic strings}",
    eprint = "2307.04691",
    archivePrefix = "arXiv",
    primaryClass = "hep-ph",
    reportNumber = "CERN-TH-2023-118, MS-TP-23-37, DESY-23-117",
    doi = "10.1088/1475-7516/2023/11/020",
    journal = "JCAP",
    volume = "11",
    pages = "020",
    year = "2023"
}

@article{Antusch:2025xrs,
    author = "Antusch, Stefan and Hinze, Kevin and Saad, Shaikh",
    title = "{Metastable cosmic strings and gravitational waves from flavor symmetry breaking}",
    eprint = "2503.05868",
    archivePrefix = "arXiv",
    primaryClass = "hep-ph",
    doi = "10.1103/528x-qzs3",
    journal = "Phys. Rev. D",
    volume = "112",
    number = "3",
    pages = "035043",
    year = "2025"
}

@article{Nielsen:1973cs,
    author = "Nielsen, Holger Bech and Olesen, P.",
    editor = "Taylor, J. C.",
    title = "{Vortex Line Models for Dual Strings}",
    doi = "10.1016/0550-3213(73)90350-7",
    journal = "Nucl. Phys. B",
    volume = "61",
    pages = "45--61",
    year = "1973"
}

@article{Gouttenoire:2019kij,
    author = "Gouttenoire, Yann and Servant, G{\'e}raldine and Simakachorn, Peera",
    title = "{Beyond the Standard Models with Cosmic Strings}",
    eprint = "1912.02569",
    archivePrefix = "arXiv",
    primaryClass = "hep-ph",
    reportNumber = "DESY-19-204",
    doi = "10.1088/1475-7516/2020/07/032",
    journal = "JCAP",
    volume = "07",
    pages = "032",
    year = "2020"
}

@article{Leblond:2009fq,
    author = "Leblond, Louis and Shlaer, Benjamin and Siemens, Xavier",
    title = "{Gravitational Waves from Broken Cosmic Strings: The Bursts and the Beads}",
    eprint = "0903.4686",
    archivePrefix = "arXiv",
    primaryClass = "astro-ph.CO",
    reportNumber = "NSF-KITP-09-36, MIFP-09-16",
    doi = "10.1103/PhysRevD.79.123519",
    journal = "Phys. Rev. D",
    volume = "79",
    pages = "123519",
    year = "2009"
}

@article{Buchmuller:2021mbb,
    author = "Buchmuller, Wilfried and Domcke, Valerie and Schmitz, Kai",
    title = "{Stochastic gravitational-wave background from metastable cosmic strings}",
    eprint = "2107.04578",
    archivePrefix = "arXiv",
    primaryClass = "hep-ph",
    reportNumber = "CERN-TH-2021-107, DESY 21-101",
    doi = "10.1088/1475-7516/2021/12/006",
    journal = "JCAP",
    volume = "12",
    number = "12",
    pages = "006",
    year = "2021"
}

@article{Buchmuller:2024zzk,
    author = "Buchmuller, Wilfried",
    title = "{Metastable strings and grand unification}",
    eprint = "2401.13333",
    archivePrefix = "arXiv",
    primaryClass = "hep-ph",
    doi = "10.22323/1.455.0004",
    journal = "PoS",
    volume = "ICPPCRubakov2023",
    pages = "004",
    year = "2024"
}

@article{Auclair:2019wcv,
    author = "Auclair, Pierre and others",
    title = "{Probing the gravitational wave background from cosmic strings with LISA}",
    eprint = "1909.00819",
    archivePrefix = "arXiv",
    primaryClass = "astro-ph.CO",
    doi = "10.1088/1475-7516/2020/04/034",
    journal = "JCAP",
    volume = "04",
    pages = "034",
    year = "2020"
}

@article{Ambjorn:1989bd,
    author = "Ambjorn, Jan and Olesen, P.",
    title = "{A Condensate Solution of the Electroweak Theory Which Interpolates Between the Broken and the Symmetric Phase}",
    reportNumber = "NBI-HE-89-26",
    doi = "10.1016/0550-3213(90)90307-Y",
    journal = "Nucl. Phys. B",
    volume = "330",
    pages = "193--204",
    year = "1990"
}

@article{Schmitz:2024gds,
    author = "Schmitz, Kai and Schroeder, Tobias",
    title = "{Gravitational waves from cosmic strings for pedestrians}",
    eprint = "2412.20907",
    archivePrefix = "arXiv",
    primaryClass = "astro-ph.CO",
    doi = "10.1088/1475-7516/2026/01/025",
    journal = "JCAP",
    volume = "01",
    pages = "025",
    year = "2026"
}

@article{Blasi:2020mfx,
    author = "Blasi, Simone and Brdar, Vedran and Schmitz, Kai",
    title = "{Has NANOGrav found first evidence for cosmic strings?}",
    eprint = "2009.06607",
    archivePrefix = "arXiv",
    primaryClass = "astro-ph.CO",
    reportNumber = "CERN-TH-2020-151",
    doi = "10.1103/PhysRevLett.126.041305",
    journal = "Phys. Rev. Lett.",
    volume = "126",
    number = "4",
    pages = "041305",
    year = "2021"
}

@article{Schmitz:2020syl,
    author = "Schmitz, Kai",
    title = "{New Sensitivity Curves for Gravitational-Wave Signals from Cosmological Phase Transitions}",
    eprint = "2002.04615",
    archivePrefix = "arXiv",
    primaryClass = "hep-ph",
    reportNumber = "CERN-TH-2020-018",
    doi = "10.1007/JHEP01(2021)097",
    journal = "JHEP",
    volume = "01",
    pages = "097",
    year = "2021"
}

@article{LIGOScientific:2025bgj,
    author = "Abac, A. G. and others",
    collaboration = "LIGO Scientific, VIRGO, KAGRA",
    title = "{Upper Limits on the Isotropic Gravitational-Wave Background from the first part of LIGO, Virgo, and KAGRA's fourth Observing Run}",
    eprint = "2508.20721",
    archivePrefix = "arXiv",
    primaryClass = "gr-qc",
    reportNumber = "LIGO-P2500349",
    month = "8",
    year = "2025"
}

@article{Tranchedone:2026lav,
    author = "Tranchedone, Lorenzo and Carragher, Ethan and Hardy, Edward and van IJcken, Nat{\'a}lie Koscelansk{\'a}",
    title = "{Metastable cosmic strings are broken at the start}",
    eprint = "2601.04320",
    archivePrefix = "arXiv",
    primaryClass = "hep-ph",
    month = "1",
    year = "2026"
}

@article{tHooft:1974kcl,
    author = "'t Hooft, Gerard",
    editor = "Taylor, J. C.",
    title = "{Magnetic Monopoles in Unified Gauge Theories}",
    reportNumber = "CERN-TH-1876",
    doi = "10.1016/0550-3213(74)90486-6",
    journal = "Nucl. Phys. B",
    volume = "79",
    pages = "276--284",
    year = "1974"
}

@article{Buchmuller:2020lbh,
    author = "Buchmuller, Wilfried and Domcke, Valerie and Schmitz, Kai",
    title = "{From NANOGrav to LIGO with metastable cosmic strings}",
    eprint = "2009.10649",
    archivePrefix = "arXiv",
    primaryClass = "astro-ph.CO",
    reportNumber = "CERN-TH-2020-157, DESY 20-154, DESY-20-154",
    doi = "10.1016/j.physletb.2020.135914",
    journal = "Phys. Lett. B",
    volume = "811",
    pages = "135914",
    year = "2020"
}

@article{Lazarides:2022jgr,
    author = "Lazarides, George and Maji, Rinku and Shafi, Qaisar",
    title = "{Gravitational waves from quasi-stable strings}",
    eprint = "2203.11204",
    archivePrefix = "arXiv",
    primaryClass = "hep-ph",
    doi = "10.1088/1475-7516/2022/08/042",
    journal = "JCAP",
    volume = "08",
    number = "08",
    pages = "042",
    year = "2022"
}

@article{Chitose:2024pmz,
    author = "Chitose, Akifumi and Ibe, Masahiro and Neda, Shunsuke and Shirai, Satoshi",
    title = "{Gravitational waves from metastable cosmic strings in supersymmetric new inflation model}",
    eprint = "2411.13299",
    archivePrefix = "arXiv",
    primaryClass = "hep-ph",
    reportNumber = "IPMU24-0043",
    doi = "10.1088/1475-7516/2025/04/010",
    journal = "JCAP",
    volume = "04",
    pages = "010",
    year = "2025"
}

@article{Chitose:2025cmt,
    author = "Chitose, Akifumi and Ibe, Masahiro and Shirai, Satoshi and Wen, Yaxuan",
    title = "{Cosmic strings in multi-step symmetry breaking}",
    eprint = "2506.15194",
    archivePrefix = "arXiv",
    primaryClass = "hep-ph",
    reportNumber = "IPMU25-0023",
    doi = "10.1007/JHEP02(2026)166",
    journal = "JHEP",
    volume = "02",
    pages = "166",
    year = "2026"
}

@article{Servant:2023tua,
    author = "Servant, G{\'e}raldine and Simakachorn, Peera",
    title = "{Ultrahigh frequency primordial gravitational waves beyond the kHz: The case of cosmic strings}",
    eprint = "2312.09281",
    archivePrefix = "arXiv",
    primaryClass = "hep-ph",
    reportNumber = "DESY-23-202, CERN-TH-2023-226",
    doi = "10.1103/PhysRevD.109.103538",
    journal = "Phys. Rev. D",
    volume = "109",
    number = "10",
    pages = "103538",
    year = "2024"
}

@article{Bian:2026tco,
    author = "Bian, Zhengyang and Chen, Ning and Guo, Mian and Hou, Zhanpeng and Ji, Haoyang and Wei, Junyi and Zhang, Zhuo",
    title = "{The non-topological $Z^\prime$ string in the 331 model and its classical stability}",
    eprint = "2604.06530",
    archivePrefix = "arXiv",
    primaryClass = "hep-ph",
    month = "4",
    year = "2026"
}

@article{Asl:2026zpj,
    author = "Asl, Doa Hashemi and Schmitz, Kai",
    title = "{New gravitational-wave templates for metastable cosmic strings: Loop breaking versus network collapse}",
    eprint = "2604.28097",
    archivePrefix = "arXiv",
    primaryClass = "hep-ph",
    reportNumber = "MS-TP-26-15",
    month = "4",
    year = "2026"
}

\end{document}